\documentclass[final,authoryear,5p,times,twocolumn]{elsarticle_accepted}

\usepackage{amsmath}

\usepackage{amssymb}

\usepackage{rotating}

\usepackage{soul}
\usepackage[usenames,dvipsnames]{color}

\usepackage{url}

\usepackage{soul}
\usepackage[usenames,dvipsnames]{color}

\usepackage{url}

\journal{Icarus}

\newcommand{\disp}{\displaystyle}

\newcommand{\REVfirst}[1]{{#1}}

\newcommand{\REVsec}[1]{{#1}}

\begin{document}
\begin{frontmatter}

\title{Dust photophoretic transport around a T Tauri star: Implications for comets composition}

\author[GSMA,UTINAM]{D.~Cordier}
\ead{daniel.cordier@univ-reims.fr}
\author[UPisa,INFN]{P.~G.~Prada Moroni}
\author[URoma,INFN]{E.~Tognelli}
\address[GSMA]{Groupe de Spectrom\'etrie Mol\'eculaire et Atmosph\'erique - UMR 6089 
               Campus Moulin de la Housse - BP 1039
               Universit\'e de Reims Champagne-Ardenne
               51687 REIMS -- France}
\address[UTINAM]{Universit{\'e} de Franche-Comt{\'e}, Institut UTINAM, CNRS/INSU, UMR 6213, 25030 Besan\c{c}on Cedex, France}
\address[UPisa]{Physics Department "E. Fermi", University of Pisa, Largo B. Pontecorvo 3, I-56127, Pisa, Italy}
\address[URoma]{Department of Physics, University of Roma Tor Vergata, Via della Ricerca Scientifica 1, 00133, Roma, Italy}
\address[INFN]{INFN, Largo B. Pontecorvo 3, I-56127, Pisa, Italy}


\begin{abstract}
  {There is a growing body of evidences for the presence of crystalline material in comets. These crystals are believed to have
   been annealed in the inner part of the proto-solar Nebula, while comets should have been formed in the outer regions. Several
   transport processes have been proposed to reconcile these two facts; among them a migration driven by photophoresis.}
  {The primarily goal of this work is to assess whether disk irradiation by a Pre-Main Sequence star would influence the photophoretic
  transport.}
  {To do so, we have implemented an evolving 1+1D model of an accretion disk, including advanced numerical techniques, 
  undergoing a time-dependent 
  irradiation, consistent with the evolution of the proto-Sun along the Pre-Main Sequence. The photophoresis is described using a formalism
  introduced in several previous works.}
  {\REVfirst{Adopting the opacity prescription used in these former studies,} we find that the disk irradiation enhances the photophoretic transport: 
  the assumption of a disk central hole of several astronomical units in radius is no longer strictly required,
  whereas the need for an {\it ad hoc} introduction of photoevaporation is reduced. 
  However, \REVfirst{we show that a residual trail of small particles could annihilate the photophoretic driven transport via their effect on 
  the opacity.} We have also confirmed that the thermal conductivity of transported
  aggregates is a crucial parameter which could limit or even suppress the photophoretic migration and generate several segregation effects.}
\end{abstract}

\end{frontmatter}


\section{Introduction}

   As reported by \cite{tielens_etal_2005} the interstellar medium (hereafter ISM) is very poor in crystalline 
solids. For instance, \cite{kemper_etal_2004} well reproduce the interstellar absorption band using 
a mixture composed of $\sim15.1$\% of amorphous pyroxene and $\sim84.9$\% of amorphous olivine by mass, leading to a crystalline 
fraction of the interstellar silicates around $0.2$\%. The proto-solar Nebula is supposed to have been formed from material 
coming from the ISM. As a consequence, the primordial dust in the Solar System should be composed of amorphous solids except grains which have undergone
either a thermal annealing in high temperature regions (\textit{i.e.} around $1000-1500$ K) close to the star (\textit{i.e.} $r \lesssim 1-2$ AU)
or a sequence starting by a vaporization or a melting and finishing by a re-condensation.
Besides this, comets are presumed to have formed in the cold outer part of the solar Nebula along the lines of the scenario recalled
below.\\
  The dissipation of the early gaseous and dusty protoplanetary disk is presumed to have left a disk of icy planetesimals beyond
$\sim 5$ AU. The Edgeworth-Kuiper Belt (hereafter KB) and the Oort Cloud (hereafter OC) are the remains of this disk and are the well
accepted two main cometary reservoirs \citep[][]{mumma_charnley_2011}. A corpus of studies, based on dynamics, show that icy asteroids
originally located between $\sim 5$ AU and $\sim 30$ AU have either participated to the formation of giant planets or have
been ejected far away because of the planet migration. The majority of the KB objects have probably been accreted at their current
distances (\textit{i.e.} $\gtrsim 30$ AU); but as suggested by results of the Nice model, an orbital resonance of Saturn and Jupiter
may have partially filled the KB with objects initially located between $\sim 15$ AU and $\sim 30$ AU \citep[][]{tsiganis_etal_2005}.
Furthermore, the planetesimals of the inner part of the primordial disk (\textit{i.e.} located in the interval $\sim 5-15$ AU)
have been likely ejected and finally participated to the formation of the outer regions of the KB and the OC. In summary, the crystals
have been incorporated into icy planetesimals when they occupied zones from $\sim 5$ AU to $\sim  30$ AU and beyond; prior to be
scattered to the KB and OC. Of course, the transport processes studied in this paper should deposit crystalline dusts at least around
$5$ AU, but the deeper the penetration beyond $5$ AU will be, the more credible the proposed mechanism will be.
    The formation of the comets is also believed to have occurred early during the solar system
formation. For instance \cite{weidenschilling_1997} showed that the formation of comets could have been completed in $\sim 2 \times 10^{5}$
years.\\
\cite{campins_ryan_1989} have found that crystalline 
olivine is a major component of the silicates in Comet Halley and \cite{wooden_etal_1999,wooden_etal_2000} detected crystalline
silicates in Hale-Bopp observations. \textit{Stardust} samples of Comet 81P/Wild 2 include large single mineral crystals
and X-ray microscopic analysis leads to a crystal mass fraction $f_{\rm cryst}$ larger than $\sim 50$\% 
\citep[][]{zolensky_etla_2006,ogliore_etal_2009,brownlee_etal_2006}. More generally, comets have a ratio $f_{\rm cryst}$ exceeding $\sim 20$\% 
\cite[][]{kelley_wooden_2009,lindsay_etal_2013}.
This discrepancy between the  
crystallinity of the ISM grains and the of cometary grains one is the mark of a radial 
transport process and/or a specific physical phenomenon occurring
in the accretion disk. Several transport processes have been proposed to explain the presence of these refractory material in comets:
annealing by shock waves in the outer solar Nebula \citep[][]{harker_desch_2002}, radial mixing by turbulent
diffusion \citep[][]{gail_2001,wehrstedt_gail_2002,bockeleeMorvan_etal_2002,cuzzi_etla_2003} or mixing in a marginally gravitationally unstable
(MGU) disk \citep[see the series of papers by Boss and co-authors: ][could be a starting point]{boss_2008}, transport by photophoresis 
\citep[][]{krauss_wurm_2005,wurm_krauss_2006,krauss_etal_2007,mousis_etal_2007,moudens_etal_2011}. 
Some authors \citep[][]{ciesla_2007,keller_gail_2004} built models including the vertical disk structure and found a radial outflow of dust grains due to pressure gradients
within the protoplanetary disk.
The X-wind mechanism, which has been advocated for the redistribution
of CAIs throughout the Nebula \citep[][]{shu_etal_1996}, could have also brought a contribution.\\
In this paper, we focus on photophoretic process for which we employ a 1+1D accretion disk undergoing a time-dependent irradiation. Indeed,
the proto-Sun, at the epoch during which it was surrounded by a disk of gas and dust, was evolving through the so-called T Tauri phase. The stars
belonging  to the T Tauri class, are known to possess an accretion disk and are evolving along the Pre-Main Sequence (hereafter PMS) tracks
where their luminosity could be much higher than 
that of the Sun. This high luminosity could have an important influence
on photophoresis and might change the disk structure or its dynamics by irradiation. This is why we were interested in modeling an accretion disk evolving consistently 
with a proto-Sun. It is of particular interest to assess if the assumption of the existence of a central gap 
\citep[as hypothesized in previous works, see ][]{moudens_etal_2011} in the disk is still required. It should be noticed that
\cite{turner_etal_2012} have taken into account the PMS luminosity of the proto-Sun in the context of Jupiter formation.\\
  The presence of crystals in comets is not the only issue for which photophoresis is supposed to play a role, 
\cite{teiser_dodsonrobinson2013} have proposed that photophoresis could accelerate the giant planets formation and
\cite{wurm_etal_2013} have investigated the involvement of this effect in the formation of Mercury-like planets and in the metal depletion in
chondrites. Detailed theoretical and experimental studies on chondrules photophoretic properties have been also conducted
 \citep[][]{wurm_etal_2010,loesche_etal_2013a}.\\
   Concerning the photophoretic transport itself we have essentially adopted the approach described in \cite{krauss_etal_2007}
and \cite{moudens_etal_2011}, as recalled in Sec.~\ref{model}. The main features of our disk model are presented in the same
section, and additional details are given in \ref{append}.
Section \ref{res} is devoted to our results, while in
Sect.~\ref{discu} we discuss some aspects of the problem such as the influence of aggregate properties, in particular that of
thermal conductivity as suggested by \cite{krauss_etal_2007}, \REVfirst{and the role of gas--dust opacity}.
Conclusions can be found in Sect.~\ref{concl}.

\section{\label{model}Description of the model}

   We have developed an original implementation of a 1+1D irradiated protoplanetary accretion disk model based on the 
equation introduced by \cite{lynden-Bell_Pringle_1974} and \cite{pringle_1981}
\begin{equation}\label{Eq_evol}
 \frac{\partial\Sigma}{\partial t} = \frac{3}{r} \frac{\partial}{\partial r} \left\{r^{1/2} 
                                     \frac{\partial}{\partial r}<\nu> \Sigma r^{1/2}\right\}
         + \dot{\Sigma}_{w}( r)
\end{equation}
  which governs the secular variations of the disk surface density $\Sigma$ (kg m$^{-2}$) as a function of the heliocentric distance $r$.
The numerical method employed to solve Eq. (\ref{Eq_evol}) is described in \ref{secevol}.
The average turbulent viscosity $<\nu>$ (m$^{2}$ s$^{-1}$), depending on a free parameter $\alpha$, is computed in the frame of the 
\cite{shakura_Sunayev_1973} formalism -- after the integration of
the vertical structure equations (see \ref{vertstruct} for details). Our disk model is based on a generalization of the method introduced by
\cite{papaloizou_terquem_1999} (hereafter PT99) and \cite{alibert_etal_2005}. The photoevaporation $\dot{\Sigma}_{w}( r)$ is provided
by the simple prescription published by \cite{veras_armitage_2004}: $\dot{\Sigma}_{w}( r)= 0$ for $r \le R_{g}$ and 
$\dot{\Sigma}_{w}( r) = \lambda_{\rm evap}/r$ for $r > R_{g}$, where $R_{g}$ is taken to be 5 AU and $\lambda_{\rm evap}$ is an adjustable
parameter. A rigorous treatment of the radiative
transfer in all its complexity is beyond the scope of the present work, and represents probably a physical and numerical challenge.
Instead, the effect of the irradiation by the central star has been taken into account by modifying the temperature boundary 
condition at the disk external surface. This way, the temperature at this surface, denoted $T_{s}$, is given by 
$T_{s}^{4}= T_{b}^{4} + T_{irr}^{4}$ where $T_{b}$ is the background temperature (usually $T_{b}= 10$ K) and $T_{irr}$ comes from 
the following equation, derived in \cite{hueso_guillot_2005},
\begin{equation}\label{Tirr}
\begin{array}{lcl}
 T_{irr} &=& T_{eff}(t)  \left[ \underbrace{\frac{\disp 2}{\disp 3\pi} \, \left( \frac{\disp R_{\star}(t)}{\disp r} \right)^{3}}_{(1)}\right. \\
         & + & \left.\underbrace{\frac{\disp 1}{\disp 2} \left(\frac{\disp R_{\star}(t)}{\disp r}\right)^{2}
            \left(\frac{\disp H_{p}(t)}{\disp r}\right)
            \left(\frac{\disp \mathrm{d}\mathrm{ln} \, H_{p}(t)}{\disp\mathrm{d}\mathrm{ln} \, r}-1\right)
            }_{(2)}
                         \right]^{1/4} \\
\end{array}
\end{equation}
where $R_{\star}(t)$ and $T_{eff}(t)$ are respectively the stellar radius and effective temperature at the age $t$; 
$H_{p}(t)$ represents the disk pressure height at the heliocentric distance $r$ and at time
$t$. This prescription has also been adopted by \cite{fouchet_etal_2012} who have emphasized its significance that we recall here:
the term (1) corresponds to the radiative flux that would be intercepted by a flat disk, the term (2) is an estimation for the effect
of the disk flaring. Following \cite{fouchet_etal_2012} and \cite{hueso_guillot_2005} we fixed
$\mathrm{d}\mathrm{ln} H_{p}/\mathrm{d}\mathrm{ln} r$ at its equilibrium value, namely $9/7$.
  In order to validate our model, we have performed several tests. For instance, we have verified that we reproduce 
the mid-plane temperature $T_{m}$, the surface density $\Sigma$ and the shape factor $H/r$ obtained by PT99 (see e.g. Figs. 2 and 3 in PT99).
We also obtained a good agreement between our computations and the PT99 ones for the accretion rate
 $\dot{M}_{st}$ (M$_{\odot}$ yr$^{-1}$) as a 
function of $\Sigma$ for various turbulence parameters $\alpha$ and distances $r$ to the Sun. In their study, PT99 did not take into
account the effect of irradiation. Notice that, this is why we verified that our computations are compatible with results found by 
\cite{dalessio_etal_1998}, even if the treatment of the irradiation is not exactly the same. 
This way, following the same procedure adopted by \cite{fouchet_etal_2012}, we also verified that  our computations are compatible with 
results found by \cite{dalessio_etal_1998}, who used a more sophisticated approach concerning irradiation.\\
  Both stars and protoplanetary disks change with time. The stars evolve under the influence of gravitation and nuclear
reactions, while disks lose their mass either by accreting on the star and/or evaporating to the interstellar space. 
Thus, we have decided to include both evolutions in a model in which the proto-Sun and the disk interact via irradiation.\\ 
  Young Main Sequence stars may exhibit a debris disk, that has a very tenuous or even non-existent gaseous component 
\citep{ollivier_etal_2009}. In contrast, T Tauri stars have their spectroscopic features (as the excess UV or IR radiation) well
explained by the presence of a surrounding, optically thick, gaseous accretion disk \citep{lynden-Bell_Pringle_1974}. The objects of this
class are Pre-Main Sequence stars, located in the Hertzsprung-Russell diagram  between the
``birthline'' and the early Main Sequence, 
where the stellar luminosity is provided by the hydrogen burning in their core with the secondary elements at equilibrium.
The ``birthline'' is a line in the 
HR diagram corresponding to a threshold below which stars become visible to the observer, they therefore begin their quasi-static contraction 
and move to the Main Sequence. We have adopted a PMS evolutionary track computed by \cite{tognelli_etal_2011} who have employed 
the well-tested and developed stellar evolutionary code FRANEC\footnote{Frascati Raphson Newton Evolutionary Code.}
\citep[][]{deglinnocenti_etal_2008}
together with up-to-date input physics \cite[][]{tognelli_etal_2012,dellomodarme_etal_2012}.\\
%
%
%
\begin{figure}[!t]
\vspace{-1cm}
\hspace{-0.5cm}\includegraphics[width=10.5 cm]{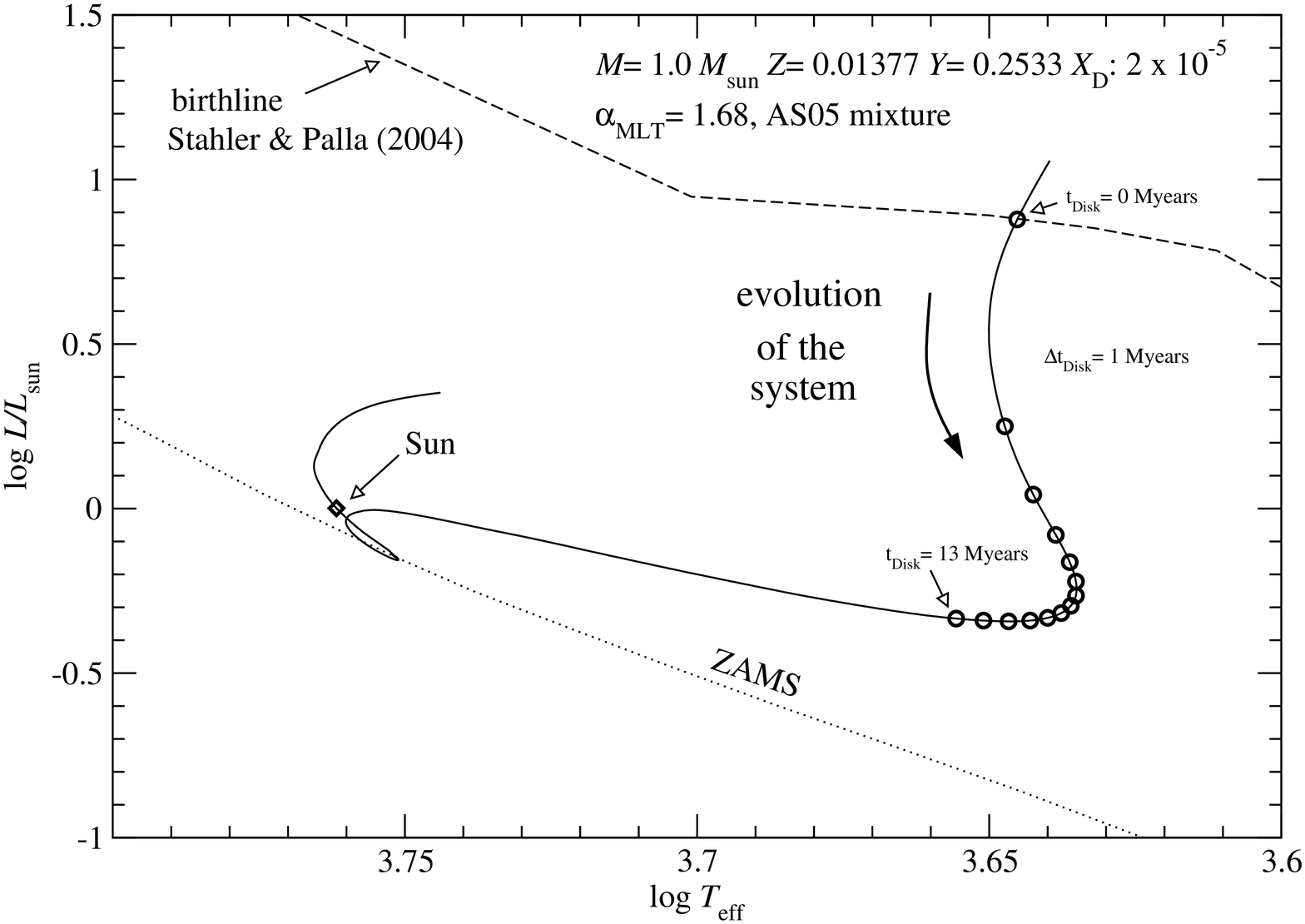}
\caption[]{\label{PMS}Solid line: Pre-Main sequence evolutionary track computed by \cite{tognelli_etal_2011}, chosen
           parameters are representative for the Sun: $Z= 0.01377$ and $Y=0.2533$ are respectively the initial heavy elements and
           the helium mass fractions, $\alpha_{\rm MLT}= 1.68$ is the solar calibrated Mixing Length Theory parameter and $X_{\rm D}$ 
           denotes the initial mass fraction of deuterium. This stellar track has been computed with the mixture of heavy elements provided
           by \cite{asplund_etal_2005} (labeled AS05).
           Dashed line: the stellar ``birthline'' from \cite{stahler_palla_2004}. 
           Dotted line: the ZAMS.
           Circles are displayed each 1 Myr and the Sun position on the track has been marked with a diamond.}
\end{figure}
%
  The starting time $t_{0}$ of the evolution of our accretion disk has been taken where the ``birthline'' 
\citep[from][]{stahler_palla_2004} intersects the proto-solar PMS stellar track.

The solar models have been calibrated on the present solar luminosity ($L_\odot$), radius ($R_\odot$), for the surface chemical composition 
($Z/X_{ph,\odot}$, see \cite{asplund_etal_2005}). We used an iterative procedure that consists in varying the initial helium abundance ($Y$), 
metallicity ($Z$), and mixing length parameter ($\alpha_{MLT}$), in order to reproduce at the age of the Sun ($t_\odot$) the present 
$L_\odot$, $R_\odot$, and $(Z/X)_{ph,\odot}$ within a tolerance of, at most, $10^{-4}$. Then, we obtained for the proto-Sun the initial 
helium and metal mass fraction $Y = 0.2533$, $Z = 0.01377$, and $\alpha_{MLT}=1.68$ (see Fig.~\ref{PMS}).  
In their study  of samples of stars belonging to six young stellar clusters, \cite{haisch_etal_2001} have measured the JHKL
 infrared excess fractions of stars. For each cluster, these ratios can be regarded as the fractions of stars
surrounded by a disk. For the selected  clusters, age determinations are available in the literature. This way,
\cite{haisch_etal_2001} have shown that, not surprisingly, the fraction of stars accompanied by a disk decreases with age and should
be negligible for an age around $\sim 6$ Myr (hereafter 1 Myr$= 10^{6}$ years). This limitation is confirmed by \cite{pascucci_tachibana_2010}
and references therein.
According to these works, protoplanetary 
disks older than $\sim 6$ Myr should not exist or be very rare. 
Our 1+1D disk model requires the pre-computation of tables of turbulent viscosity $<\nu>$ depending on various parameters among which
the luminosity of the star $L_{\star}$ and its effective temperature $T_{eff}$. 
Consequently, along the PMS track we took a set of couples $(L_{\star},T_{eff})$ corresponding to ages $t \lesssim 16$ Myr
with $t=0$ at the ``birthline''. Hence, for each $t$, tables of mean viscosity were constructed using $(L_{\star}(t), T_{eff}(t))$ or
equivalently $(R_{\star}(t),T_{eff}(t))$ and playing with a set of parameters values: $\alpha= 10^{-3}$ and $10^{-2}$; $\dot{M}_{st}$ ranging
from $10^{-12} \, M_{\odot}$ yr$^{-1}$ to $10^{-4}\, M_{\odot}$ yr$^{-1}$; and $0.05 \le r\le 50$ AU. A dedicated subroutine
allows interpolation in mentioned tables of $<\nu>_{t,\Sigma,r}$.\\
   The dust grains transported through the disk are aggregates of small particles 
\citep[][]{brownlee_1978,greenberg_1985,meakin_donn_1988,blum_etal_2000}{ } and are 
very approximately considered as spherules of radius $a$. Hereafter, the terms ``particles'', ``grains'' and ``aggregates'' will be
synonyms; while ``monomer'' will be reserved for the components of aggregates.\\

  Following the approach developed by \cite{krauss_wurm_2005} and \cite{krauss_etal_2007} we assume that the gas flow conditions are
described by the Knudsen number, $K_{n}$, which is defined as $K_{n}= l/a$, where $l$ is the mean free path of the gas molecules.\\
  An expression of the photophoretic force $F_{ph}$, valid for free molecular regime ($K_{n} > 1$) and for continuum regime
($K_{n} \le 1$), has been proposed by \cite{beresnev_etal_1993} (see their Eq. (31))
\begin{multline}
\label{phforce}
 F_{ph} =  \frac{\pi}{3} a^{2} I(r,t) J_{1} \left(\frac{\pi m_{g}}{2 kT}\right)^{1/2} \, \\
           \frac{\alpha_{E} \psi_{1}}{\alpha_{E} + 15 \Lambda Kn (1-\alpha_{E})/4 + \alpha_{E} \Lambda \psi_{2}} 
\end{multline}
 We are not going to recall the expression of each term of Eq. (\ref{phforce}), instead we invite the interested reader to consult
previous works \citep[][]{beresnev_etal_1993,krauss_etal_2007,moudens_etal_2011}. Nonetheless, we specify some important points.
First of all, given that the collision/scattering cross section of molecules is not known, the free mean path $l$ is estimated 
from the value of the dynamic viscosity $\eta$ provided by $\eta= \eta_{0} \sqrt{T/T_{0}}$
\citep[see][]{krauss_etal_2007} with $\eta_{0}= 8.4 \times 10^{-6}$ Pa s, $T$ the temperature and $T_{0}= 280$ K.
Since the viscosity $\eta$ for a dilute gas is also given by $\eta= n m \bar{v} l /3$ \citep[with $n$ the number of molecules
per unit of volume and $m$ the average mass of these molecules,][]{reif_1967}, we compute 
$l= 3\eta /\rho_{m} \bar{v}$ where $\rho_{m}$ is the mid-plane density and $\bar{v}$ the thermal velocity
of gas molecules.\\
  The parameter $\Lambda$ that appears in Eq. (\ref{phforce}) measures the thermal relaxation of aggregates, we have 
\citep[see][]{krauss_etal_2007} $\Lambda= \lambda_{eff}/\lambda_{g}$, with $\lambda_{g}$ the thermal conductivity of gas and 
$\lambda_{eff}$ the effective thermal conductivity given by the following expression,
\begin{equation}
\lambda_{eff} = \lambda_{p} + 4 \epsilon \sigma T^{3} a
\end{equation}
 where $\epsilon$ is the emissivity (assumed to be 1) and $\lambda_{p}$ is the heat conductivity of aggregates. As usual,
$\sigma$ denotes the Stefan-Boltzmann constant. Although minerals, in massive and not porous form, have thermal conductivity around
$\sim 10$ W.m$^{-1}$ K$^{-1}$ \citep[see for instance][]{horai_simmons_1969}, 
%
%
\begin{figure}[!t]
\includegraphics[width=9 cm]{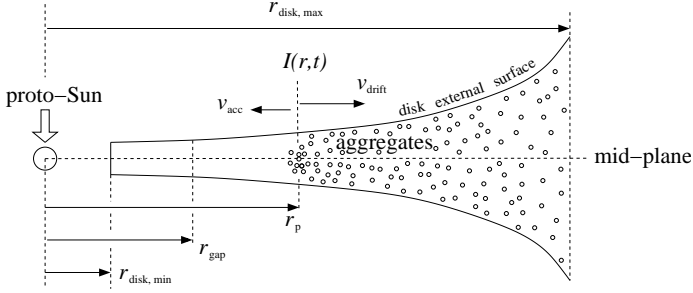}
\caption[]{\label{sketch}Sketch of the photophoretic transport model discussed in this paper. The disk structure has been computed
           between $r_{disk, min}= 0.5$ AU and $r_{disk,max}= 50$ AU. The existence of an inner gap of radius $r_{gap} \ge r_{disk,min}$
           is hypothesized.}
\end{figure}
%
aggregates -- due to their porosity -- are believed worse
thermal conductors than massive dust grains. In their experimental investigations of laboratory analogs, \cite{krause_etal_2011}
found thermal conductivity ranging from $0.002$ to $0.02$ W m$^{-1}$ K$^{-1}$. We have fixed $\lambda_{p}$ to $10^{-3}$
W m$^{-1}$ K$^{-1}$ in order to facilitate comparisons with earlier researches having adopted this value \citep[][]{mousis_etal_2007,moudens_etal_2011,krauss_wurm_2005}.
Nevertheless, in Sect.~\ref{discu} we will explore the influence of $\lambda_{p}$. 
The density of dust particles have been fixed to $\rho_{\rm p}=500$ kg m$^{-3}$; this value has been adopted by \cite{mousis_etal_2007}
and \cite{moudens_etal_2011}. 
This density value is based on an olivine density of $3300$ kg m$^{-3}$ \citep{mousis_etal_2007}, and
on an aggregates filling factor of $15$\% \citep[][]{blum_schrapler_2004}. The influence of $\rho_{\rm p}$ is briefly discussed in
Sect.~\ref{influ_rhop}.
  The photophoretic force (see Eq. \ref{phforce}) is not the only one taken into account; the force $F_{res}$ due to residual gravity
\citep[][]{weidenschilling_1977} and the radiative pressure force are also considered. These forces are respectively provided by
\begin{equation}\label{Fres}
F_{res}= \frac{\disp m_{p}}{\rho_{g}} \, \frac{\disp\mathrm{d}P}{\disp\mathrm{d}r}
\end{equation}
\begin{equation}\label{Frad}
F_{rad}= \pi a^{2} \frac{\disp I(r,t)}{\disp c_{\rm light}}
\end{equation}
  where $m_{p}$ is the mass of an aggregate, $\rho_{g}$ and $P$ are the density and the pressure of the gas; $I(r,t)$ represents
the radiative flux (W.m$^{-2}$) at time $t$, at the location of the test particle. The speed of light is noted $c_{\rm light}$.
    The radiative flux $I(r,t)$, as it appears in Eq. (\ref{phforce}) and (\ref{Frad}) is derived from the star luminosity
$L_{\star}(t)$. 
Similarly to the previous works \citep[][]{mousis_etal_2007,moudens_etal_2011}, we have hypothesized the existence of an 
inner gap of radius $r_{\rm gap}$, optically thin enough for particles to see the proto-Sun, but still containing the gas content of 
the disk structure. This assumption is supported by an increasing number of observational evidences. For instance, \cite{dalessio_etal_2005}
show that observations of the Pre-Main Sequence star CoKu Tau/4 suggest the presence of an accretion disk with an inner hole cleared
of small dust grains. The infrared imaging survey, conducted by \cite{siciliaaguilar_etal_2006} with the Spitzer Space Telescope,
indicates that around 10\% of the stars of their sample, owning a disk, exhibit spectral features explained by the existence of an optically thin inner
disk. In addition, \cite{pontoppidan_etal_2008} studied three disks, and detected an inner gap depleted in dust particles but containing
gas. Up to now, it is not clear which physical mechanism could be the origin of such inner cavities or gaps in protoplanetary disks
\citep[see for instance][]{thalmann_etal_2010}. This
is the reason why we have only postulated their existence.
The model is sketched in Fig. \ref{sketch}. The radiative transfer along the line of sight between particles and proto-Sun is treated
in a very simplified way. Firstly, neglecting the absorption by the gas, the radiative flux $I_{geom}$ (W m$^{-2}$) is evaluated 
at the distance $r$ reached by particles at the time $t$
\begin{equation}
 I_{\rm geom}(r,t)= \frac{\disp L_{\star}(t)}{\disp 4\pi r^{2}}
\end{equation}
  this reflects the sphericity of stellar emissions. In a second step, the equation
\begin{equation}
 \mathrm{d}I= -\kappa_{R} \, \rho_{m}(r,t) \, I(r,t) \, \mathrm{d}r
\end{equation}
 is integrated at fixed $t$ from $r_{gap}$ to $r_{p}$ using the boundary condition $I(r_{gap},t)= I_{\rm geom}(r_{p},t)$; in this manner
the effect of gas absorption is combined with the pure geometrical decrease. The opacity $\kappa_{R}$ is due
to Rayleigh's scattering for which \cite{mousis_etal_2007} derived
\begin{equation}\label{kappaR}
\kappa_{R}(T_{eff})= 3.96 \times 10^{-19} \, T_{eff}^{4}(t)
\end{equation}
(cm$^{2}$.g$^{-1}$). It has to be emphasize that, using Eq. (\ref{kappaR}), the disk is assumed to have been cleared of dust by the
photophoretic transport between $r_{\rm gap}$ and $r_{p}(t)$ (see Fig.~\ref{sketch}). We also point out that the employment of the opacity law provided
by \cite{bell_lin_1994} (see \ref{vertstruct}) is not more relevant, since it assumes the absence of dust depletion caused by the 
photophoretic transport. With an overestimated contribution of dust grains, \cite{bell_lin_1994} opacity law yields to the
annihilation of the photoretic migration. Thus, the approach adopted here is clearly an idealized situation and the derived results have to be 
considered as maximized effects. \REVfirst{We will come back to this opacity issue in a dedicated section of the discussion.}
The use of Eq. (\ref{kappaR}) means that the effect of the thermal radiation field of the
gas itself and the effect of the photon multi-scattering have also been neglected.
Finally, the aggregates are drifted (in the gas frame) in the radial direction with the velocity (see Fig. \ref{sketch})
\begin{equation}\label{Vdrift}
v_{\rm drift}= \frac{\disp F_{\rm ph} + F_{\rm rad} + F_{\rm res}}{\disp m_{p}} \, \tau
\end{equation}
 where $\tau$, the coupling time of particles with gas, is provided by 
\begin{equation}\label{tau}
 \tau= \frac{\disp m_{p}}{\disp 6\pi \eta a} \, C_{c}
\end{equation}
 The correction factor $C_{c}$ is expressed as a function of the Knudsen number $K_{n}$ \citep[][]{cunningham_1910,hutchins_etal_1995}
\begin{equation}\label{Cc}
 C_{c}= 1 + K_{n} \, (1.231 + 0.47 \, \mathrm{e}^{-1.178/K_{n}})
\end{equation}
  In protoplanetary disks, the mid-plane temperature decreases with the distance to the central star. This is why we have checked that
the thermophoretic force given by \cite{young_2011}, and recalled in \cite{lutro_2012}, remains negligible compared to the forces
contributing to Eq. (\ref{Vdrift}).
 For a given age of the disk, the distance $r_{p}$ reached by particles is obtained by integrating the difference between the drift
velocity $v_{\rm drift}$ and the accretion velocity $v_{\rm acc}$, we have
\begin{equation}\label{rp}
r_{p}= \int_{0}^{\rm age} (v_{\rm drift}-v_{\rm acc}) \, \mathrm{d}t
\end{equation}
  More rigourously than \cite{moudens_etal_2011}, we derived $v_{\rm acc}$ from the disk model, that provides naturally
the speed of the gas which is falling to the star (see Fig. \ref{sketch})
\begin{equation}\label{vacc}
v_{\rm acc}= - \frac{\disp 3}{\disp \Sigma \sqrt{r}} \, \frac{\disp\partial}{\disp\partial r}\left(\Sigma <\nu> \sqrt{r}\right)
\end{equation}
\citep[][]{lynden-Bell_Pringle_1974}.

\section{\label{res}Results}

\begin{figure*}[!t]
\hspace{-0.50cm}\includegraphics[angle=-90, width=20cm]{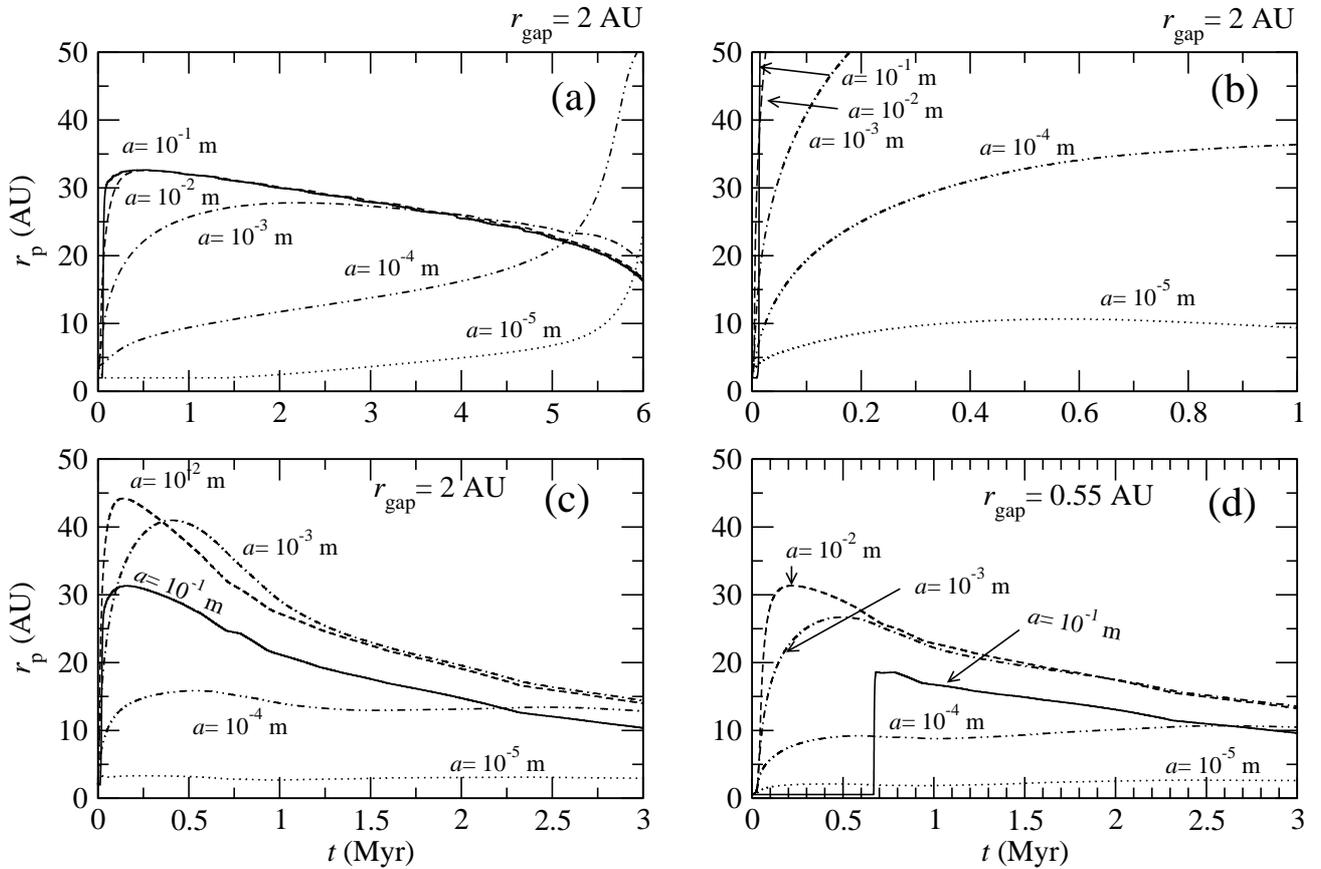}
\caption[]{\label{firstresults} In this figure we have displayed the heliocentric distance $r_{p}$ (AU) reached by aggregates 
           of differents radius $a$ ranging from $10^{-5}$ m to $10^{-1}$ m, as a function of time $t$ (Myr). In all represented 
           computations, the initial mass of the simulated proto-solar Nebula has been fixed to 3 MMSN and the turbulent viscosity 
           parameter is $\alpha= 7 \times 10^{-3}$.
           (a) The disk structure is not irradiated and the Sun luminosity is constant, fixed at the ZAMS value. An inner gap with
           a radius of 2 AU is assumed to occupy the central part of the disk. 
           (b) The disk is still not irradiated, the Nebula structure is the same as in panel (a). However, time-dependent 
           effective temperature $T_{eff}(t)$ and the luminosity $L_{\star}(t)$ have been used for the photophoretic and radiative forces
           estimations. 
           (c) The disk evolution is simulated taking into account the irradiation by the star evolving along its PMS track, 
           the forces acting upon the particles are computed consistently, an inner gap of $2$ AU is still assumed.
           (d) The same simulation as the one represented in panel (c) but the hypothesis of the existence of an inner gap has been
           abandoned; the inner gap radius of $0.55$ AU corresponds pratically to a very reduced gap.
           }
\end{figure*}

\begin{figure}[!t]
\vspace{-1.5cm}
\hspace{-0.5cm}\includegraphics[angle=0, width=11cm]{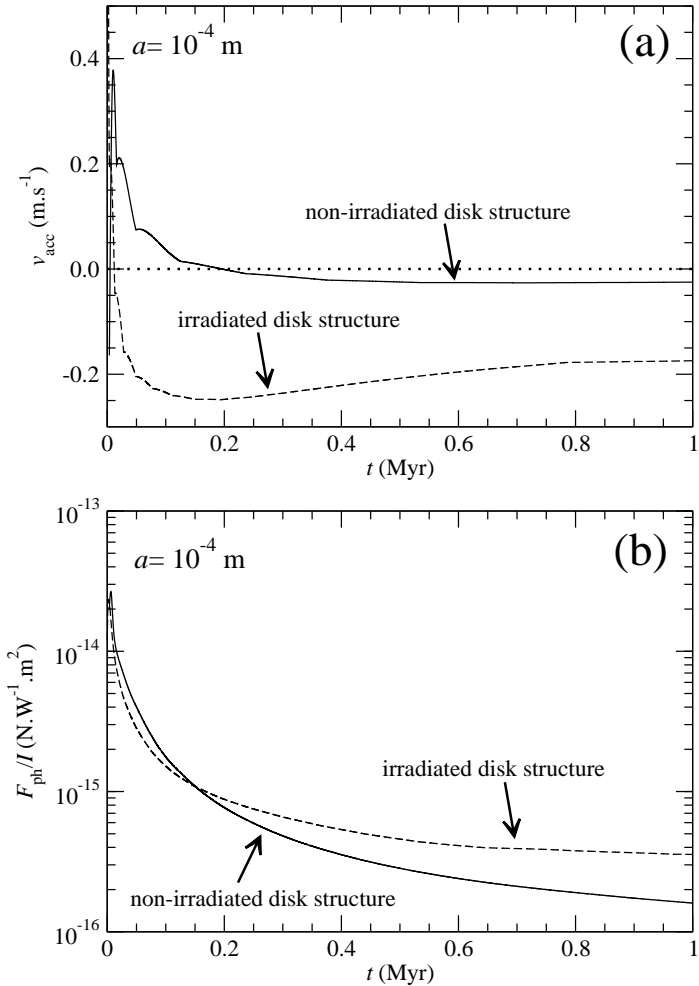}
\caption[]{\label{expla}(a) The gas accretion velocity $v_{\rm acc}$ (see Eq.~\ref{vacc}) undergone by a dust particle along its
           ``trajectory'' through the disk mid-plane. Two cases are distinguished: a non-irradiated disk (solid line),
           an irradiated disk (dashed line). 
           (b) The photophoretic force $F_{ph}$ divided by the radiative flux $I$ as it is seen by a test particle during its
           migration in the Nebula. 
           Both examples correspond to a particle size of $a= 10^{-4}$ m, a Nebula initial mass of $3$ MMSN and a turbulent 
           viscosity parameter $\alpha= 7 \times 10^{-3}$.}
\end{figure}

  The concept of the Minimum Mass of solar Nebula (hereafter MMSN) dates back to the end of the seventies \citep[][]{weidenschilling_1977b,hayashi_1981},
and the initial total mass of the solar Nebula is now expected to be approximately a few MMSN 
\citep[\textit{e.g.}][found $\sim 10$ MMSN]{crida_2009}.
We have chosen $3$ MMSN as a typical value, that allows easy comparison with past investigations \citep[][]{moudens_etal_2011}.
   The proto-planetary disks are believed to be the place of turbulent mixing produced by magneto-rotational instabilities
\cite[][]{balbus_hawley_1998} and the associated $\alpha$-parameter introduced by \cite{shakura_Sunayev_1973} has admitted values
within the interval $10^{-2}-10^{-3}$; we have selected $\alpha= 7 \times 10^{-3}$ as our nominal value which also facilitates 
comparisons. The literature reported the observation of accretion disks having a central hole with a radius of $\sim 1-2$ AU
\cite[see for instance][]{besla_wu_2007,pontoppidan_etal_2008,hughes_etal_2010,thalmann_etal_2010}; for this reason we first 
assumed the existence of such an inner gap
of radius $r_{\rm gap}= 2$ AU. This strong hypothesis, already made by \cite{mousis_etal_2007} and \cite{moudens_etal_2011}, favors clearly
the transport driven by photophoresis since it decreases the radiative flux attenuation between the proto-Sun and the dust grains.\\
The results of our first calculations can be seen in Fig.~\ref{firstresults}(a), they are very similar to those
plotted in Fig. 4 of \cite{moudens_etal_2011}; for which equivalent parameters and hypothesis have been chosen. We recall that, in this case,
the lifetime of the protoplanetary disk has been adjusted to $6$ Myr \citep[see][]{haisch_etal_2001,pascucci_tachibana_2010} by tuning the photoevaporation
parameter $\lambda_{\rm evap}$. In Moudens \textit{et al.}'s work the disk is not irradiated, and the solar luminosity was provided by \cite{pietrinferni_etal_2004}
for the Sun at the ZAMS\footnote{Zero Age Main Sequence.}. For each simulation of a disk evolution, we stopped the calculation when
the total mass of the disk reached $\sim 1$\% of its initial mass.
This criterion will
be applied in all simulations discussed in the rest of the present work.\\
In Fig.~\ref{firstresults}(b) an identical disk
structure is kept but the photophoresis force is computed in a slightly different way: the solar luminosity is no longer taken 
constant at its ZAMS value, but instead it follows the luminosity variations along the PMS tracks plotted in Fig.~\ref{PMS}. 
In addition,
the gas opacity which depends on the stellar effective temperature $T_{eff}$ (see Eq. \ref{kappaR}), is computed consistently. 
Not surprisingly, the aggregates are very efficiently pushed outwards, 
because the luminosity at the birthline is about $10$ times larger than that at the ZAMS.
Interestingly, particles with a radius of $a= 10^{-5}$ m reach 
$\sim 10$ AU after $\sim 0.3$ Myr and do not move further away even after $6$ Myr.\\
  Keeping fixed the inner gap at $2$ AU, and letting unchanged all the other parameters, we adopted a protoplanetary
%
disk model computed with a time-dependent stellar luminosity given by the already mentioned PMS track. The 
subsequent positions of the aggregates are represented in Fig.~\ref{firstresults} (c). It is clearly visible that the use of an 
irradiated disk damps the particle movement. This behavior could be caused either by a higher accretion velocity or by smaller forces.
In order to disentangle the various effects, we have focused our analysis on the movement of particles of radius $10^{-4}$ m, 
which appear to be typical at the examination of Fig.~\ref{firstresults} (b) and Fig.~\ref{firstresults} (c). The panel (a)
of Fig.~\ref{expla} indicates that the accretion velocity undergone by the considered particle, along its path in the disk mid-plane,
is larger (in absolute value) when the disk structure is irradiated. Beside this, we have checked that during the period of
interest, \textit{i.e.} before $\sim 1.5$ Myr, the photophoretic force has the dominant contribution to the drift velocity (see Eq.~\ref{Vdrift}).
To catch the effect of the disk structure, we have normalized the photophoretic force by the received flux of light (see Eq. \ref{phforce}).
The resulting ratio $F_{ph}/I$ is either not significantly different or slightly higher,
respectively when irradiation is accounted or when it is not (see Fig.~\ref{expla}.b). Therefore, the larger accretion velocity  
appears to weaken the efficiency of the photophoretic force during the T Tauri phase of the Sun.
%
\begin{table*}[t!]
\caption[]{Lifetimes (in Myr) of the disk model as a function of the initial mass (in MMSN) and the turbulent mixing parameter $\alpha$.
           All these simulations were performed without photoevaporation but they include the time-dependent irradiation by the 
           PMS Sun.}
\begin{center}
{
\begin{tabular}{lcccccc}
\hline
                &           &                     &   $\alpha$         &                    &                    &           \\
\hline
Mass  (in MMSN) & $10^{-3}$ &  $2 \times 10^{-3}$ & $4 \times 10^{-3}$ & $6 \times 10^{-3}$ & $8 \times 10^{-3}$ & $10^{-2}$ \\
\hline
   1            & 16.4      &   10.7              & 7.9                &  5.8               & 4.0                & 3.0       \\
   3            & 15.7      &   10.4              & 7.7                &  5.3               & 3.7                & 2.8       \\ 
  10            & 15.1      &   10.0              & 7.4                &  4.8               & 3.3                & 2.5       \\ 
\hline
\end{tabular}
}
\end{center}
\label{dissiptime}
\end{table*}
%
We emphasized that this irradiated disk model is dissipated at the age of $4.32$ Myr. Therefore, the photoevaporation has not been needed 
to get a lifetime compatible with astronomical observations. Indeed, with non-irradiated disks the photoevaporation rate was
adjusted to get a disk lifetime of $6$ Myr. With an initial mass of $3$ MMSN, $\alpha= 7 \times 10^{-3}$ and no 
photoevaporation, the disk is dissipated in $12.6$ Myr. A photoevaporation rate of 
$\dot{M}_{w}= 1.425 \times 10^{-9} \; M_{\odot}$ yr$^{-1}$ is therefore required  
%
%
%
to reduce the dissipation time to $6$ Myr. 
 In fact, we found that for $r \gtrsim 15$ AU, irradiation increases significantly (\textit{i.e.} by a factor up to $\sim 4$) the mean
turbulent viscosity
$<\nu>=\int_{-H}^{+H}\rho \nu \mathrm{d}z/\Sigma$, where $\nu$ is computed in the frame of the \cite{shakura_Sunayev_1973}'s
formalism: $\nu= \alpha c_{s}^{2}/\Omega$ ($\Omega$ being the Kepler's frequency and $c_{s}$ the speed of sound). For an ideal gas:
$c_{s}^{2}= P/\rho \propto T$; at relatively large distances for the star (\textit{i.e.} beyond $\sim 15-20$ AU), the radiative flux
rises the temperature in the thickness of the disk, leading to the observed increase in $<\nu>$, and consequently to the shorter
dissipation times. We stress that this fact is consistent with the larger observed $v_{acc}$
(see Fig.~\ref{expla}.a).
Table~\ref{dissiptime} summarizes the obtained lifetimes for initial masses of $1$, $3$ and $10$ MMSN, and $\alpha$ parameter
ranging between  $10^{-3}$ and $10^{-2}$. As we can see, for simulations corresponding to $\alpha$ higher than $\sim 6 \times 10^{-3}$
the lifetimes are compatible with observations. Obviously, this does not mean that photoevaporation 
does not exist \citep[see for instance][]{owen_2006}{ } but rather that our improved models are in better agreement with astronomical
observations, and do not required systematically large assumed photoevaporation rates.
   As it can be noticed in Tab~\ref{dissiptime}, the computed lifetimes seem to depend slightly on the initial mass, this behavior
is due to the chosen criterion for stopping the simulation ({\it i.e.} the total mass decreased to $1$\% of its initial value): for the largest
initial masses the mass accretion rates are the highest causing a criterion satisfied earlier.
  Finally, the inner gap radius has been reduced from $2$ AU to $0.55$ AU. In order to ease comparisons with previous works, all
our reference disk structures have been computed between $0.50$ and $50$ AU \cite[][ determined the edge of the KB 
around $\sim 50$ AU]{allen_etal_2001}. The inner boundary of $0.55$ AU has been chosen instead of $0.50$ AU so that to avoid numerical difficulties 
caused by the finite-differences 
%
\begin{table}[htbp]
\caption[]{Maximum heliocentric distances (in AU) reached by aggregates of radius $a$ ranging between $10^{-1}$ and $10^{-5}$ m.
           The disk is irradiated and $\alpha= 7 \times 10^{-3}$.}
\begin{center}
{\small
\begin{tabular}{lcccccc}
$a$ (m)         & $10^{-1}$        &  $10^{-2}$      & $10^{-3}$      &  $10^{-4}$      &   $10^{-5}$   \\
\hline
1 MMSN          & 15               &  27.5           & 28             &  23             &   50          \\
\hline
3 MMSN          & 19               &  32             & 27             &  10             &   5           \\
\hline   
10 MMSN         & 17               &  30.5           & 19             &   8             &   2           \\ 
\hline
\end{tabular}
}
\end{center}
\label{influMMSN}
\end{table}
%
calculation of derivatives with respect to $r$. As it could be
expected (see Fig.~\ref{firstresults}. d), the distances $r_{p}$ are in average lower than those computed
in the previous case where $r_{\rm gap}= 2$ AU, this is clearly the consequence of a stronger light absorption along the line of sight:
since the difference $r_{p}-r_{\rm gap}$ (see Fig.~\ref{sketch}) is larger, the flux of photons received by a given aggregate becomes fainter, 
thus producing weaker photophoretic and radiative forces. Remarkably, dust particles with radius larger than $10^{-4}$ m can be found up to 
$20$--$30$ AU from the proto-Sun and consequently could participate to the formation of icy planetesimals within the KB.\\
  In Tab.~\ref{influMMSN} we have gathered the maximum distances reached by dust aggregates when the initial mass of the Nebula ranges
between 1 MMSN and 10 MMSN. Clearly, a low mass Nebula favors large distances for small particles. We would like to stress 
that this is not only because small particles ($\sim 10^{-4}-10^{-5}$ m) get to outer regions that larger aggregates could not be 
found in cometary material:
indeed aggregation/coagulation processes could form relatively big grains using small ones as building blocks
\citep[][]{guttler_etal_2010,zsom_etal_2010,blum_etal_2014}. 
Alternatively, the opposite process could occur: fragmentation could produce small particles from relatively large ones.\\
We have also performed irradiated
disk simulations with $\alpha= 10^{-2}$ and $\alpha= 10^{-3}$; limiting --when necessary-- the age to $6$ Myr, by adjusting the 
photoevaporation parameter. The results were essentially similar to those obtained at $\alpha= 7 \times 10^{-3}$ (displayed in
Tab.~\ref{influMMSN}).\\
    The inner limit of the disk structure (\textit{i.e.} the above mentioned $0.5$ AU) may appear somewhat arbitrary. Nonetheless, even
in the case of T Tauri stars which have not developed an extended inner hole (\textit{i.e.} up to distances of several AU) the accretion
disk is truncated in the vicinity of the star. Indeed, T Tauri stars are known to be magnetically very active; \textit{e.g.}, Zeeman 
broadening measurements have revealed surface field in the range of $1-3$ kG. Such strong magnetic fields are at the origin of ``jets''
that eject material to the interstellar medium, and also produce a magnetospheric accretion in which the disk matter is channeled
onto the star along magnetic field lines \cite[][]{ferreira_etal_2006,bouvier_etal_2007} out of the mid-plane, leaving it optically
thin. Although, this quantity is the subject of star-to-star substantial variations, the radius of this magnetospheric accretion zone
can be reasonably estimated to be around $\sim 0.1$ AU. We, then, computed a specific set of evolving
disk structures using $r_{\rm gap}= 0.10$ AU. Keeping other parameters unchanged compared to those used for the simulations reported
in Fig.~\ref{firstresults}(d), we obtained only aggregates with radius of $a= 10^{-3}$ m and $a= 10^{-2}$ m reaching heliocentric distances
slightly larger than $\sim 6$ AU while the other dust particles remain closer to the star within $5$ AU. These outputs shows, one more time, that 
the existence of an inner hole, optically thin and having an expansion beyond $\sim 0.5$ AU, is a determining factor for the efficiency 
of the photophoretic transport of dust in protoplanetary disks.
%
\section{\label{discu}Discussion}
%
\subsection{The Kinetics of the Grains Annealing in an Irradiated Disk}
  The problem of the existence of crystals in comets raises naturally the question of the kinetics of thermal annealing of silicate
grains that originate from the interstellar medium. Indeed, this lattice structure transformation cannot be instantaneous; in addition, one
can wonder if grains at relatively low temperature could be annealed in timescale compatible with disk lifetime. If the process could
occur at large heliocentric distances with a duration $\lesssim 6$ Myr, then the transport of grain from regions in the
vicinity of the proto-Sun, to larger distances would be no longer needed. \cite{lenzuni_etal_1995} gave the typical duration $t$
for converting an amorphous domain to a crystalline domain: $t = \nu^{-1} \, \mathrm{e}^{D/k_{B} T_{a}}$, where $\nu$ is the 
characteristic vibrational frequency of silicate, $D$ is the activation energy of repositioning atoms within the lattice structure,
$T_{a}$ the temperature and $k_{B}$ the Boltzmann's constant. \cite{lenzuni_etal_1995} gave also the
typical values $\nu \sim 2.5 \times 10^{13}$ s$^{-1}$ and $D/k_{B} \sim 41000$ K, which leads to $t \sim 5$ Myr for $T_{a}= 645$ K and
$23$ Myr for $T_{a}= 630$ K. We have checked in our models of irradiated disks that the mid-plane temperature $T_{m}$ remains always
below $\sim 100$ K disk life-long for distances larger than a few AU. As a consequence, amorphous grains cannot be annealed, by gas temperature,
at distances compatible with comets
formation during the disk lifetime, even if the disk is irradiated. Concerning the production of crystals by vaporization/recondensation,
the temperatures required to vaporized silicates ({\it i.e.} $T \le 1500$ K) are never reached in our models in the external Solar System.
\REVfirst{As a consequence, the remaining possibility is an evaporation/condensation sequence in high temperature regions --{\it i.e.} those
close to the proto-Sun-- during the early phases of the disk evolution.}
%
%
\subsection{\label{influ_rhop}The influence of monomers density}
   In this work, each dust particle is an agglomerate of smaller grains. In real accretion disks, these ``elementary grains'' may
show a distribution in size, shape and composition. Here, like in previous ones, the aggregates have been represented by
a radius ({\it i.e}. a typical size), a thermal conductivity (whose influence is discussed in Sect.~\ref{influ_conducti}) and a density
that determines the mass of each aggregate. This density is a function of the assumed aggregate porosity and of the average density
of monomers. \cite{blum_schrapler_2004} conducted laboratory experiments consisting of random ballistic deposition of monodisperse 
SiO$_2$ spheres with $1.5$ $\mu$m diameter and found an --already mentioned-- volume filling factor of $0.15$. More recently,
\cite{zsom_etal_2010} using sophisticated Monte-Carlo simulations, based on a detailed modelization of collisions, derived filling 
factors around $0.30$. This leads to a higher dust particle density, of the order of $1000$ kg m$^{-3}$ rather than $500$ kg m$^{-3}$
previously employed. By replacing $500$ kg m$^{-3}$ with $1000$ kg m$^{-3}$, for a non-irradiated model of an intial mass of $3$ MMSN
and evolving for $6$ Myr (this model is comparable to the one plotted in Fig.~\ref{firstresults}.(a) \REVfirst{which has been computed taking $500$
kg m$^{-3}$}) we obtained a maximum distance reached by aggregates that decreases, due to the larger inertia of particles. 
Not surprinsingly, a dependence with respect to the dust particle size \REVfirst{has been found}: 
the larger the particle is (\textit{i.e.} large radius $a$), and the lower is the maximum distance reached. For instance, for 
radii $a$ in the range $10^{-1}$--$10^{-2}$ m the distance reduction is of the order of $15$\%, while it falls to $\sim 10$\% for $a= 10^{-3}$
m, and becomes negligible for $a= 10^{-5}$ m. Compared to the other sources of uncertainties, particularly those regarding the
actual thermal conductivity of aggregates, the effect of dust \REVfirst{monomer density does not appear to be the dominant one ({\it i.e.} when replacing
our fiducial value of $500$ kg m$^{-3}$ by $1000$ kg m$^{-3}$)}.
%
%
\subsection{\label{influ_conducti}The influence of thermal conductivity and porosity of aggregates}

\begin{figure}[!t]
\vspace{-0.65cm}
\hspace{-0.5cm}\includegraphics[angle=0, width=11cm]{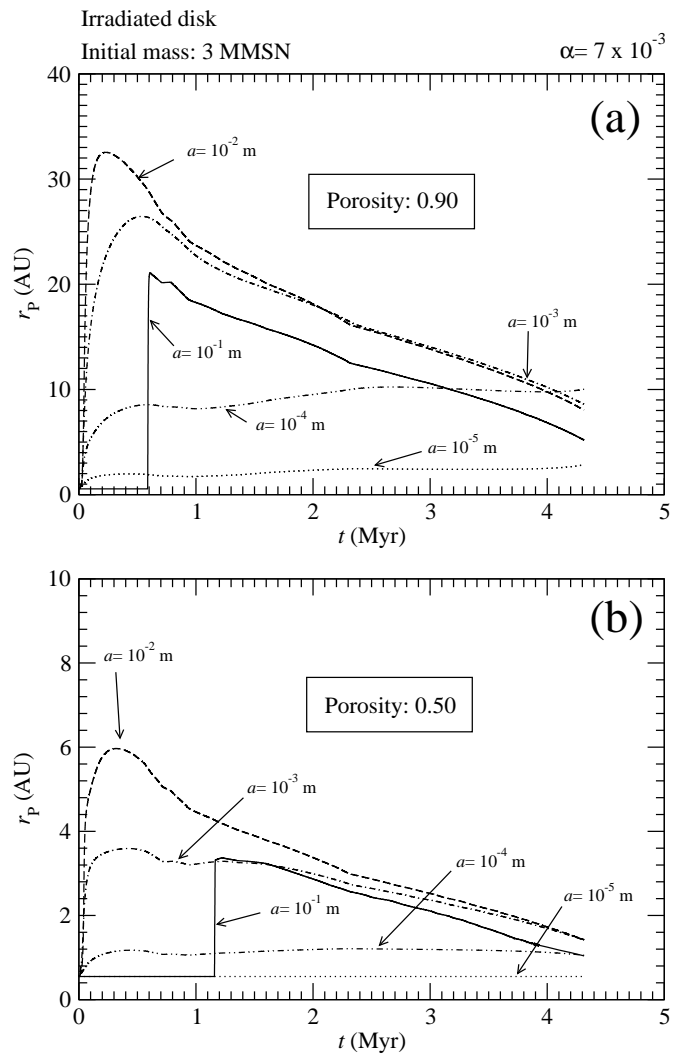}
\caption[]{\label{influ_Conduct_Density}(a) Heliocentric distances $r_{p}$ reached by particles of various sizes $a$, in an
           irradiated disk similar to one used in Fig.~\ref{firstresults}(d); the thermal conductivity has been given by
           Eq. (\ref{EqKrause11}) for a porosity $\Pi= 0.90$, the density of aggregates has been estimated consistently.
           (b) The same as in panel (a), but taking $\Pi= 0.50$.
           }
\end{figure}

     The thermal conductivity of aggregates $\lambda_{p}$ remains poorly constrained, and 
could depend on many parameters: the temperature, the exact nature of the bulk material, the porosity, the shape and the size distribution of the
monomers, the number and the size of contact areas between monomers, the lattice structure of the monomers, etc. The heat
conductivity is a crucial quantity in the context
of photophoresis: high values should diminish the effect of photophoresis because they facilitate the uniformization of the
temperature over the ``surface'' of each particle. In contrast to this, if aggregates are bad thermal conductors the photophoretic
effect is expected to be very efficient. 
    So far, we used the value $\lambda_{p}= 10^{-3}$ W m$^{-1}$ K$^{-1}$ which is the same 
adopted by \cite{mousis_etal_2007} and \cite{moudens_etal_2011}; although this is an extreme value, it has been adopted by these authors 
without any well referenced justification. Surprisingly, \cite{krauss_etal_2007} 
\citep[who have several authors in common with][]{mousis_etal_2007,moudens_etal_2011} have worked with $\lambda_{p}= 10^{-2}$ W m$^{-1}$ K$^{-1}$
and have suggested that the precise influence of $\lambda_{p}$ should be explored. The value used by \cite{krauss_etal_2007} 
comes from \cite{presley_christensen_1997} who have not reported value as low as $10^{-3}$ W m$^{-1}$ K$^{-1}$ for relevant materials.
In addition, \cite{vonBorstel_blum_2012}, who found thermal conductivities around $10^{-1}$ W m$^{-1}$ K$^{-1}$, mentioned that
results published by \cite{moudens_etal_2011} should be quantitatively affected by a higher heat conductivity. 
For all these reasons, detailed investigations about the importance of thermal conductivity 
are needed.\\
%
%
%
   \cite{krause_etal_2011} conducted laboratory experiments with aggregate analogs composed by monodisperse spherical monomers.
Their samples of 1.5 $\mu$m-sized SiO$_2$ particles were prepared following several specific protocoles which led to various values
of the porosity. This way, \cite{krause_etal_2011} obtained a range of thermal conductivity that lies between
$0.002$--$0.02$ W m$^{-1}$ K$^{-1}$. \cite{gundlach_blum_2012}, in their work on the heat transport in porous 
surface dust layers, found values compatible with such a range. 
We should notice that the lowest value of \cite{gundlach_blum_2012}, 
\textit{i.e.} $0.002$ W m$^{-1}$ K$^{-1}$, is close the one that we have used. For an irradiated disk
model of $3$ MMSN computed with $\alpha= 7 \times 10^{-3}$, the use of $\lambda_{p}= 0.02$ W m$^{-1}$ K$^{-1}$ yields to a 
maximum heliocentric distance reached by particles around $10$ AU (corresponding to $a= 10^{-2}$ m). This result has to be compared
to what it is depicted in Fig.~\ref{firstresults}.d and thus illustrates the high sensitivity of photophoresis with respect to thermal
conductivity.\\
  \cite{krause_etal_2011} were able to derive an empirical law providing the thermal conductivity of their analogs as a function of 
the volume filling factor, or equivalently of the porosity $\Pi$. This relationship can be expressed as
\begin{equation}\label{EqKrause11}
\lambda_{p} = 0.000514 \, \exp(7.91 \, (1 - \Pi))
\end{equation}
(in W m$^{-1}$ K$^{-1}$). Straightforwardly, the density of the composite material can be written 
$\rho_{p} = (1 - \Pi) \, \rho_{\rm olivine} + \Pi \, \rho_{\rm gas}$, where $\rho_{\rm olivine}$ represents the density of monomers
of non-porous olivine, we have found $\rho_{\rm olivine}\sim 4 \times 10^{3}$ kg m$^{-3}$ \citep[][]{kogel_etal_2006}, $\rho_{\rm gas}$ 
is the local density of the gas. In summary, Eq. (\ref{EqKrause11}) together with the above mentioned expression of $\rho_{p}$ offers the possibility of an
exploration of the combined effects of thermal conductivity and of the density, for several values of the porosity. 
Fig.~\ref{influ_Conduct_Density} displays results of two of such simulations performed respectively with $\Pi= 0.90$ and $\Pi= 0.50$.
The work of \cite{zsom_etal_2010}, who found $\Pi\sim 0.60$ in the mid-plan at $1$ AU, favors the latter value.
With the set of physical inputs involved there, {\it i.e} an irradiated disk computed from an initial mass of $3$ MMSN, using
a turbulent viscosity parameter $\alpha= 7 \times 10^{-3}$ and with a small ({\it i.e.} $0.55$ AU in radius) assumed inner gap, the aggregates with $\Pi= 90$\% 
are transported, to probable comets formation zone, but a relatively strong gradient in porosity is obtained. Only the smallest
aggregates did not exceed $5$ AU (see Fig.~\ref{influ_Conduct_Density}.a). In this frame, a ``mild porosity'' around $50$\% 
should annihilate any substantial photophoretic driven migration (see Fig.~\ref{influ_Conduct_Density}.b). Of course, if one reintroduces a
hypothesized larger inner gap, optically thin, a lower porosity would be permitted.
%
%
%
%
  In actual accretion disk, the ambient gas fills the empty spaces within the aggregates. The thermal conductivity of the gas, which
depends on the local thermodynamic conditions, should affect the effective conductivity of the whole aggregate. Because the presence
of gases would have contributed to the measured conductivity, \cite{krause_etal_2011} have operated under high vacuum conditions, setting
the pressure within their chamber around $10^{-5}$ mbar (\textit{i.e.} around $10^{-3}$ Pa). Then, their measurements have not to be
taken at face, but rather have to be corrected by the effect of the gas incorporated in the porous structures.
  In the context of ceramic materials, \cite{russell_1935} has derived the effective thermal conductivity of a dry porous material
taking into account the properties of its component gas and of the solid. A uniform distribution of pores in a cubic lattice and
a parallel heat flow are assumed, the convection across the pores is neglected. Russell found
\begin{equation}\label{lambda_porosity}
  \frac{\disp \lambda_{s}}{\disp \lambda_{p}} = 
        1 - \Pi^{1/3} + \frac{\disp \Pi^{1/3}}{\disp (\lambda_{\rm gas}/\lambda_{s}) \Pi^{2/3} + 1 - \Pi^{2/3}}
\end{equation}
In this formalism $\lambda_{p}$ is the conductivity of the composite material (here the aggregate), $\lambda_{\rm gas}$ 
and $\lambda_{s}$ respectively the gas conductivity and the one of the solid (here olivine). 
  Unfortunately, we found that Eq. (\ref{lambda_porosity}) is not in agreement with \cite{krause_etal_2011} results, even when the
contribution of gas is neglected. Nevertheless, we have computed the gas heat conductivity along the particles tracks, we obtained
values around a few $10^{-2}$ W m$^{-1}$ K$^{-1}$, as a consequence the effective conductivity of aggregates for porosity as high as
$90$\% should be close to $\sim 10^{-2}$ W m$^{-1}$ K$^{-1}$ and the photophoretic migration through the disk mid-plane 
should be considerably reduced.\\
   Another aspect of the issue discussed here relates to the thermal conductivity of the bulk material of the dust particles. 
\cite{krause_etal_2011} have employed amorphous silica in their experiments. In the case of solid vitreous SiO$_2$, the heat 
conductivity remains in the interval $0.85$--$1.30$ W m$^{-1}$ K$^{-1}$ depending on the temperature \citep[][]{Handbook_2005}. 
However, actual aggregates are very likely mainly made of olivine. 
The thermal conductivity of olivine could be estimated by mean of equation (12)
published in the article by \cite{xu_etal_2004} and parameters values provided therein. By doing this, we found a conductivity
around $3.7$ W m$^{-1}$ K$^{-1}$. Thus, aggregates made of olivine could be expected to exhibit a higher heat conductivity than
those built from amorphous monomers of SiO$_2$.\\
     All these arguments, together with the properties of 
the turbulence do not support far migrations of dust particles, but favor transport by photophoresis limited to maximum distances between $\sim 5$ AU and $\sim 30$ AU 
accompanied by several segregation processes on porosity, chemical nature and size; processes which will be difficult to disentangle.\\
   The emissivity $\epsilon$ is also a thermodynamic parameter. We have checked that it has no noticeable influence on the 
maximum heliocentric distances reached by grains. Decreasing the value of  $\epsilon$ (we made test with $\epsilon= 0.5$ and
$\epsilon= 0.1$) increases by a tiny amount the efficiency of photophoresis.\\
   If thermal effects reduce or impede dust migration driven by photophoresis in the disk mid-plane, we could examine whether more
favorable conditions may be found within other zones of the Nebula. This is what it will be discussed in the Sect.~\ref{offplane}.
%
%
%
\subsection{\label{influ_opacity}The Influence of the Opacity}
%
\begin{figure}[!h]
\includegraphics[angle=-90, width=9cm]{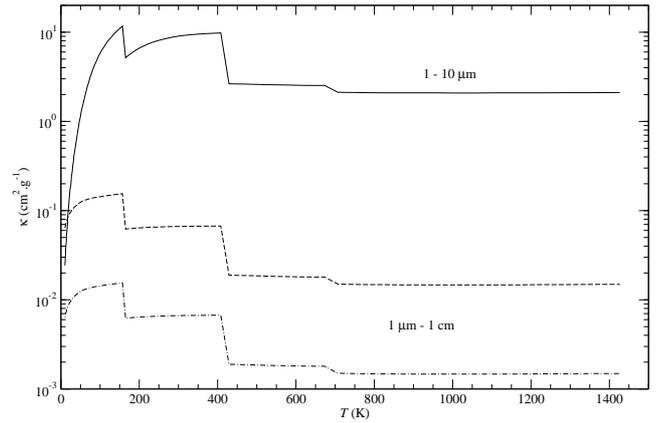}
\caption[]{\label{opacity_cuzzi_etal2014}\REVfirst{The opacity $\kappa$ of a gas--dust mixture relevant for a protoplanetary disk
           \citep{cuzzi_etal_2014}. A dust-to-gas mass ratio of $0.014$ and a power law distribution in size are assumed for all curves. Solid line:
           the range of sizes spans from $1$ to $10$ $\mu$m, dashed line: the interval of radii is $1$ $\mu$m--$1$ cm. The dash-dotted
           curve corresponds to a  $1$ $\mu$m--$1$ cm but with a total mass fraction of solids divided by ten compared to $0.014$.}}
\end{figure}
%
\REVfirst{As it can be seen in Fig.~\ref{firstresults}, regardless of considered scenario, the velocity of migration of ``test particles''
strongly depends on their size. The tendency is that small particles move slower than bigger ones. For the computation of the trajectories,
the employed prescription --also used in previous published works-- assumes the existence of the dust-free space between the inner boundary
of the disk and the position of the ``test particles'' $r_p$ (see Fig.~\ref{sketch}). This simplifying hypothesis allows the use of the opacity
law given by Eq.~\ref{kappaR} that accounts only for the gas Rayleigh's scattering, which is valid in the absence of dust grains. In observed
accretion disk, as in the ISM, the populations of grains are not monodisperse but rather follow a size distribution
\citep[see for instance][and references therein]{apai_lauretta_2010}. As it has been noticed, the transport of the largest grains is faster than
that of small particles, then some amount of small aggregates should remain along the line of sight between the Sun and the largest dust grains.
Unfortunately, the small grains are the most numerous as the distribution of the grains size is believed to follow a powerlaw $a^{-n}$ with
$n \sim 3$, and consequently they produce a dominant contribution to the opacity. Indeed, in Fig.~\ref{opacity_cuzzi_etal2014} one can compare the opacity laws
\citep{cuzzi_etal_2014}
resulting from an aggregates sizes distribution spanning from $1$ to $10$ $\mu$m and another with radii between $1$ $\mu$m--$1$ cm. Clearly, 
the population dominated by the smallest particles produces the highest opacity.\\
  For our purpose, the relevant distribution in size is the one with radii between $1$ $\mu$m--$1$ cm. As we can see in Fig.~\ref{opacity_cuzzi_etal2014}
the corresponding opacity lies between $10^{-2}$ and $2\times 10^{-1}$ cm$^2$ g$^{-1}$. These values have to be compared with an estimation made
using Eq.~\ref{kappaR}. Using this latter formula, we obtained an opacity of the order of $\sim 1.4 \times 10^{-4}$ cm$^2$ g$^{-1}$ for an effective temperature of the
proto-Sun around $4400$ K. It demonstrates that the opacity of the gas alone is order of magnitudes lower than what we find if a residual amount
of dust is left. As a conclusion, if large particles (for instance with a radius $a \sim 1$ cm) are pushed away by some transport mechanism, then
the small ones remaining on the optical path have an effect on the opacity strong enough to annihilate any force produced by photophoresis and/or
radiation pressure. Hence, two alternatives would be possible: (1) all the grains are displaced at the same velocity, which is that of the smallest
aggregates, or (2) the biggest particles are transported over a fraction of AU, the residual population of smaller grains coagulate and restablish
a distribution of sizes containing a fraction of large dusts but with a global mass of solids disminished. The process could then repeat and extend itself, tending to a
dust-free gas in the disk mid-plane.\\
   Concerning the first scenario, the comets formation zone could be provided in crystal only if the smallest aggregates reach a far enough distance
during the lifetime of the protoplanetary disk. In our approach, this could occur in the case of a non-irradiated disk (see Fig.~\ref{firstresults}.a)
which is not the most realistic model. Unfortunately, when the irradiation is taken into account, particles with radius as small as $10^{-5}$ m seem to 
stay at few AU from the Sun. In this situation, the second scenario could be invoked. In that case, the coagulation process has to be fast enough;
such a scenario looks plausible since \cite{ormel_etal_2007} found that aggregates can grow to radii up to $\sim 10$ cm in a few thousand years. The depletion in
solids created locally by the short-range migration of the biggest grains yields to a more transparent medium by decreasing the opacity
(see dash-dotted curve in Fig.~\ref{opacity_cuzzi_etal2014}). The proposed mechanism, which differs from that described in studies
like \cite{krauss_wurm_2005} --who did not take into account the feedback on opacity-- has to be studied in details in future researches.
In-depth investigations of this scenario could decide if the net effect lets the grains piling up at the inner edge of the disk or if this solid material
could be efficiently swept out from inner regions. The answer is not straightforward. 
For instance, we have noticed that a high porosity of aggregates (which favors the photophoresis by producing low thermal conductivities) is
an unfavorable factor by enhancing the opacity. In addition, to crown it all, the thermal conductivity of grains seems to depend on their size 
\citep[see][]{presley_christensen_1997}. Beside these effects, in future investigations, one has to include coagulation/fragmentation and 
the detailed evolution of the aggregates population over time, together with their influence on the opacity, has to be consistently followed.}\\
  \REVsec{Throughout this paragraph, the physical processes were assumed to take place in the disk plane of symmetry, which is supposed to embed
the majority of dust grains. Of course, the mid-plane of the disk is also the optically thickest region and the material becomes progressively 
optically thinner and thinner moving outwards.
For instance, we have estimated the optical depth $\tau_{10}$ at $\sim 10$ AU at the ``external surface'', and we have made comparison
with its mid-plane counterpart. Using \citep{cuzzi_etal_2014} opacities (that include the effect of dust) we got $\tau_{10}$ ranging between $\sim 35$ 
and more than $100$ at the mid-plane, while $\tau_{10}$ never exceeded $\sim 0.6$ during the same disk lifetime of $6$ Myr, when evaluated at
the ``external surface''. Then, even if the disk
mid-plane stays very opaque, transport induced by stellar photon flux could occur through other optically thinner regions.}
%
%
\subsection{\label{offplane}About a possible off mid-plane transport}
   As mentioned above, we have mainly assumed that particles were transported through the disk mid-plane. In real accretion disks,
dust particles are not confined in the mid-plane, but they rather explore the entire thickness of the disk between the 
mid-plane and the external surface (see Fig.~\ref{sketch}). As the surface is less dense and 
cooler compared to that at the mid-plane, we can expect quite different drift velocities. In fact, a less 
dense matter is more transparent and thus favors a larger effect of the radiation pressure. Concerning the photophoretic force, the 
net effect is more uncertain: a less opaque gas should cause an easier heating of particles (that favors the photophoretic effect, if
this heating is restricted to one ``side'' of each particle) 
while the scarcity of molecules available for photophoresis would be a limiting factor. Throughout this paragraph, one has to keep in mind
that the dust grains are probably much more abundant in the disk mid-plane than at the external surface, for instance in the steady-state
regime the vertical distribution of grains follows a gaussian law \cite[see for instance][]{birnstiel_2011}. Nonetheless, the determination
of the actual vertical distribution in our context is, by far, beyond the primarily scope of the present paper.\\
   Concerning the transport of particles at the external surface of the disk, one can argue that 
irradiation by the stellar wind and/or by energetic galactic cosmic rays, could damage their lattice and 
subsequently produces an irradiation-induced amorphization
(see for instance \cite{fama_etal_2010} for a discussion concerning the water ice or \cite{leguillou_etal_2013} for an experimental
study of the effect of electron irradiation of kerogens). Nevertheless, the small cross-section of dust particles could prevent this
effect. In future works, a more in depth discussion is needed about this aspect.\\
%
%
\begin{figure*}[!t]
\hspace{0cm}\includegraphics[angle=-90, width=18cm]{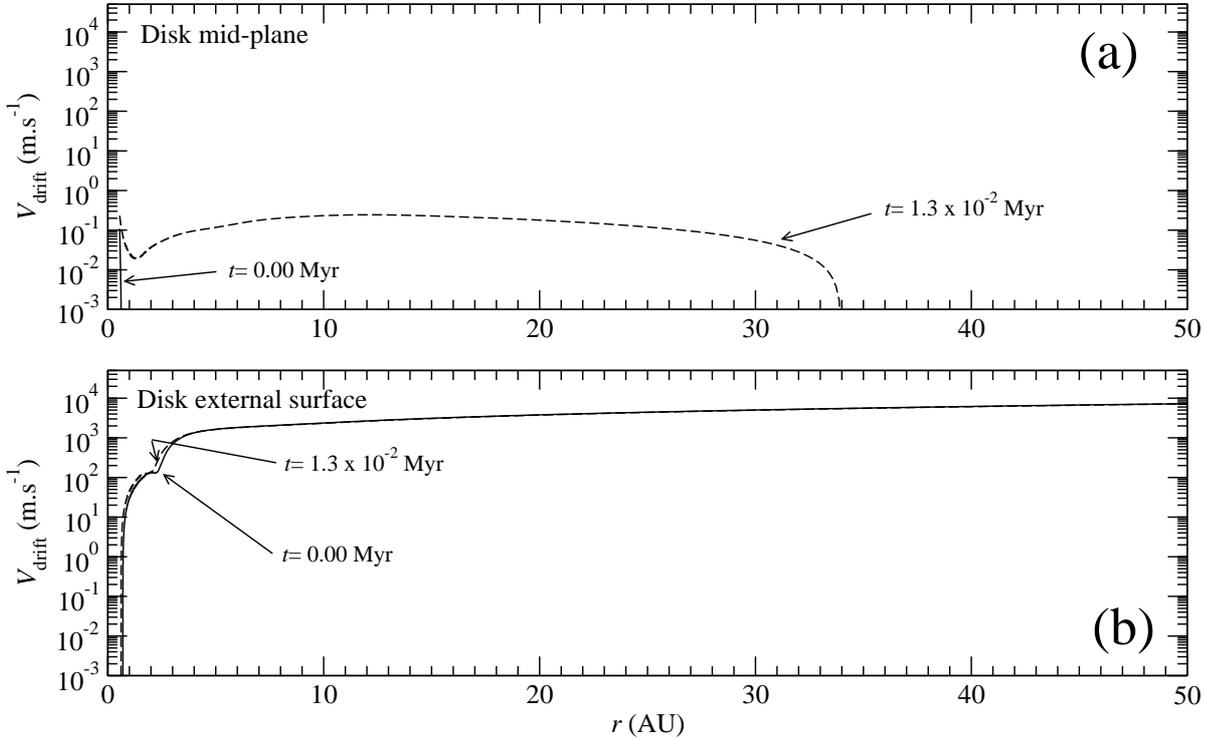}
\caption[]{\label{Vdrift}The drift velocity computed: (a) in the disk mid-plane, (b) at the external surface, for the initial
           disk model (\textit{i.e.} at $t= 0.00$ Myr) represented by a solid line and at $t= 1.3 \times 10^{-2}$ Myr (corresponding to
           a disk mass of about $\sim 90$\% of the initial mass) shown by a dashed line, in this case the lines are almost merged. 
           All these computation were done assuming a particle radius of $10^{-3}$ m. For these tests the adopted thermal conductivity
           is our ``standard'' value, {\it i.e}. $10^{-3}$ W m$^{-1}$ K$^{-1}$.}
\end{figure*}
%
      For the sake of simplicity, the radiation attenuation
will be computed as if the external surface was flat. The adopted disk model is a 3 MMSN irradiated disk, computed with 
$\alpha= 7 \times 10^{-3}$.
 Concerning the computation of the photophoretic force we keep our ``standard'' value for the involved parameters; particularly,
the thermal conductivity $\lambda_{\rm p}$ is set to $10^{-3}$ W m$^{-1}$ K$^{-1}$. As the accretion velocity $v_{\rm acc}$
(see Eq.~\ref{vacc}) is a vertically averaged quantity, we did not compute the trajectory of aggregates by using Eq.~\ref{rp} that
requires a local expression of $v_{\rm acc}$, relevant for the considered surface. We then calculated only the drift velocity
$v_{\rm drift}$ and found very high values. They are much higher than $v_{\rm acc}$ (to be considered as a typical value) and even
higher than the speed of sound. This large velocity makes no physical sense by itself, this is the reason why we have searched what
caused such very high values.
%
%
\begin{figure*}[!t]
\hspace{0cm}\includegraphics[angle=-90, width=19.5cm]{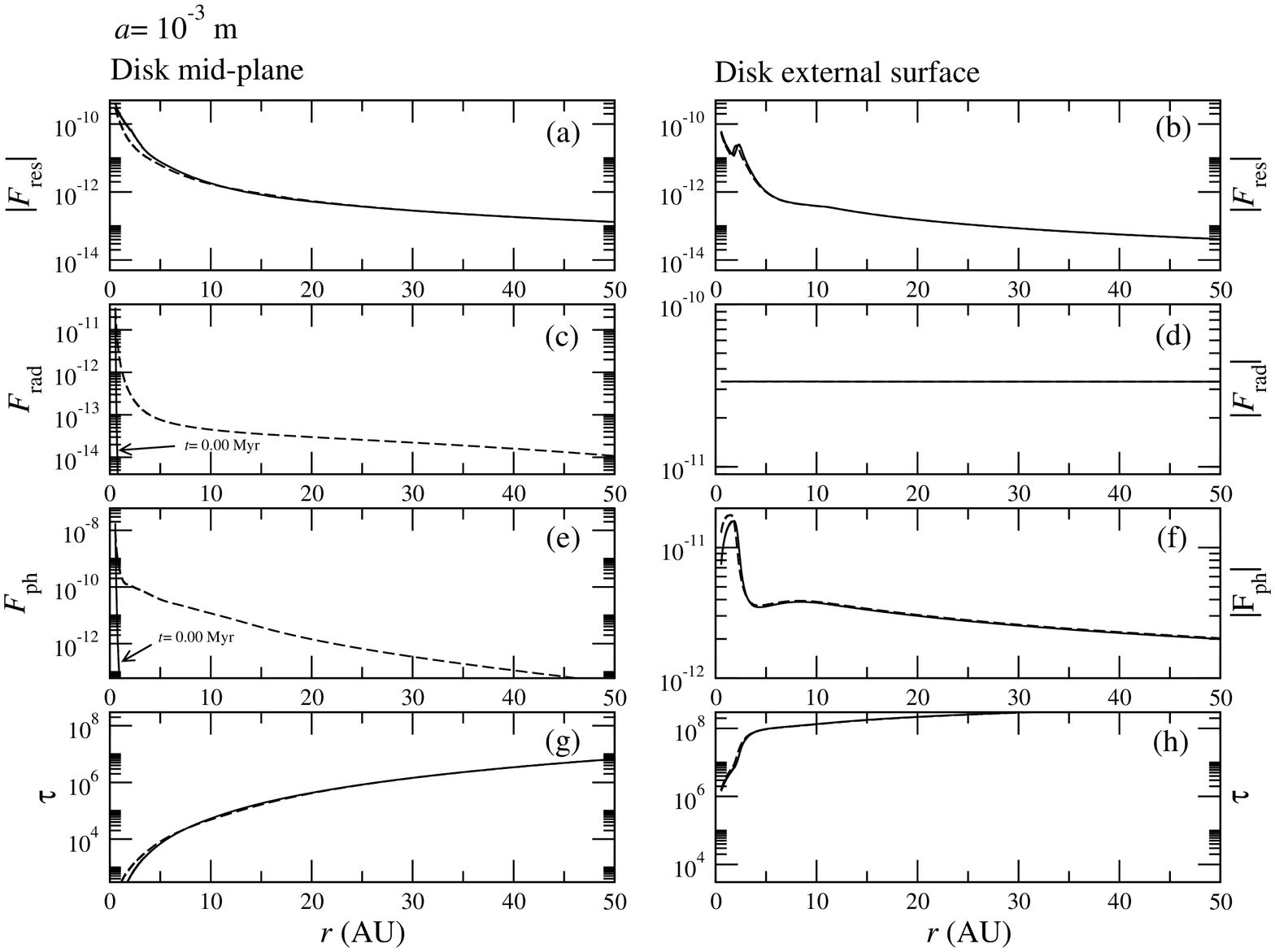}
\caption[]{\label{Fphetc}Comparison of forces and gas--coupling time at mid-plane and at the external surface. In all
           panels a solid line has been used to represent quantities related to $t= 0.0$ Myr while a dashed one has been 
           drawn for $t= 1.3 \times 10^{-2}$ Myr. The assumed particle radius is in all cases $10^{-3}$ m. In all panels, except
           in (c) and (e) the two curves merge each other. For these computations we kept the thermal conductivity of
           aggregates monomers at our ``standard'' value, {\it i.e}. $10^{-3}$ W m$^{-1}$ K$^{-1}$.}
\end{figure*}
%
     To do so, we have computed the quantities contributing to $v_{\rm drift}$ (\textit{i.e.} $F_{\rm res}$, $F_{\rm rad}$, etc)
at all points 
along the upper surface layer; this, for two disk ages of the early evolution. All these quantities were computed using the local pressure
and temperature (\textit{i.e.} those at the external surface).
  In Fig.~\ref{Vdrift} we have represented the velocity $v_{\rm drift}$ for the adopted 
initial disk model (corresponding age: 0.00 Myr) and at the time when about $10$\% of the initial mass has been accreted 
(\textit{i.e.} age: $\simeq 1.3 \times 10^{-2}$ Myr). The choice of these two ages is relevant for our purpose since they will bracket
the early evolution of the system.
The aggregate radius $a$ has been fixed to $10^{-3}$ m for these simulations. As can be seen in
Fig.~\ref{Vdrift}, at $t= 1.3 \times 10^{-2}$ Myr and $r\sim 1$ AU the velocity $v_{\rm drift}$ at the external surface (see Fig.~\ref{Vdrift}b) is
roughly $4$ orders of magnitude larger than the corresponding value 
at the mid-plane (see Fig.~\ref{Vdrift}a). 
  To go further, we have plotted the forces $|F_{\rm res}|$ (in general $F_{res} < 0$ because $P$ decreases when $r$ increases), 
$F_{\rm rad}$ and $F_{\rm ph}$ (see respectively Eq.~\ref{Fres}, \ref{Frad} 
and \ref{phforce}) together with the coupling time of particles with gas (see Eq.~\ref{tau}) in Fig.~\ref{Fphetc}. The same
disk model is employed, all these quantities have been computed respectively at the external surface (right hand side of Fig.~\ref{Fphetc}) 
and at the
disk mid-plane (left hand side of Fig.~\ref{Fphetc}); and respectively for $t= 0.0$ Myr and $t= 1.3 \times 10^{-2}$ Myr. It can be noticed that
at mid-plane the force of residual gravity $|F_{\rm res}|$ is the largest at the beginning of the disk evolution. After a few times, it
turns out that the photophoretic force $F_{\rm ph}$ dominates other contributions.
%
\begin{table*}[t!]
\caption[]{Forces (in newtons) acting on a particle of radius $10^{-3}$ m located at $2$ AU from the central star. 
           The notation $x.yz \times 10^{-n}= x.yz \, (-n)$ is
           used. The times $t_{0}$ and $t_{1}$ are respectively the intial time and $t= 1.3 \times 10^{-2}$ Myr.}
\begin{center}
{\small
\begin{tabular}{lcc|cc|cc}
\hline
                & $|F_{\rm res}|$ &                 & $F_{\rm rad}$   &                 & $F_{\rm ph}$    &                 \\
                & $t_{0}$         & $t_{1}$         & $t_{0}$         & $t_{1}$         & $t_{0}$         & $t_{1}$         \\
\hline
mid-plane       & $6.8 (-11)$ & $3.0 \, (-11)$ & $1.3 \, (-18)$ & $3.2 \, (-13)$ & $1.6 \, (-16)$ & $1.1 \, (-10)$ \\
ext. surface    & $1.6 (-11)$ & $2.3 \, (-11)$ & $3.3 \, (-11)$ & $3.4 \, (-11)$ & $1.5 \, (-11)$ & $1.2 \, (-11)$ \\  
\hline
\end{tabular}
}
\end{center}
\label{forcesat2AU}
\end{table*}
%
     At the external surface, the force due to the residual gravity is lower than at the disk mid-plane; all the three forces have a
similar order of magnitude (\textit{i.e.} $\sim 10^{-11}$ N) although the radiative pressure force dominates slightly the others. The major difference
between situations at external surface and mid-plane consists in the role played by $F_{\rm rad}$ at the surface. 
   Tab.~\ref{forcesat2AU} gives precise value of forces at $2$ AU from the proto-Sun.
We can remark that the
photophoretic force does not undergo a large change between the two locations: the effect of the higher radiative flux at the
surface seems to be compensated by the increasing scarcity of gas molecules contributing to the photophoresis. Finally, the
global resulting force (\textit{i.e.} given by the sum $F_{\rm res}+F_{\rm rad}+F_{\rm ph}$) appears to be not so different at disk mid-plane
and at external surface (\textit{i.e.} it ranges between $\sim 10^{-11}-10^{-10}$ for our test particle). 
As a consequence, the quantitatively significant difference between mid-plane and external surface comes from factor of $10^{4}$ in
gas--grain coupling time $\tau$ (see Fig.\ref{Fphetc} panels (g) and (h) for $r\sim 2$ AU). This ratio of 4 orders of magnitude is
due to local thermodynamic conditions.
  We recall that $\tau \propto C_{c}/\eta$ (see Eqs. \ref{tau} and Eq. \ref{Cc}) in a low density environment.
Moreover, in low density regions  $\eta$ becomes very low implying large $1/\eta$ values, while the Knudsen's number $K_{n}$ tends 
to be very large, and for $K_{n} \gg 1$
we have $C_{c} \simeq 1.701 \, K_{n}$, so that $C_{c}$ also increases. The net result is a strong rise of $\tau$. The
physical meaning of these large value of $\tau$ is a weak coupling between the aggregates and the gas.\\
  In summary, at the external boundary of the dense region of the protoplanetary disk the photophoretic force may not be the dominant
force, particularly if the thermal conductivity $\lambda_{\rm p}$ is much higher than $10^{-3}$ W m$^{-1}$ K$^{-1}$. The fact that
the aggregates particles are weakly coupled to the gas allows any small force to generate a transport process. In such circumstances,
particles can sediment, be pushed outward by stellar wind bursts, taken away by photoevaporation, etc. A study of such transport
processes is far beyond the scope of the present paper and required a minimum modelization of the disk's ``atmosphere''.
%
%
\section{\label{concl}Conclusion and perspectives}

   The 1+1D disk model used here implements the classical $\alpha$-viscosity prescription that allows a description of the physical
evolution of the disk as it undergoes mass and angular momentum transport. To this standard approach, we have added time-dependent
irradiation which is consistent with the evolution of the proto-Sun along the Pre-Main Sequence phase. Since this phase corresponds to
the T Tauri period of our star, which is known to be contemporary of the existence of a circumstellar dusty and gaseous accretion
disk, our model represents a noticeable improvement of the proto-solar Nebula modeling. This is particularly true in our context, in which
we examine the influence of proto-Sun radiation on the transport of dust driven by photophoresis.\\
  \REVfirst{By adopting the same prescription used in previous works for the opacity of the gas}, we have found that the high luminosity of 
the Sun considered in its T Tauri phase favors the migration of dust
grains to the outer parts of the protoplanetary disk. However, the effect is not as high as if the irradiation would not change the
disk properties. Indeed, the irradiation enhances the computed turbulent viscosity which in turn increases the accretion velocity.
Consequently, the particles are slowed down whereas the disk lifetime is reduced. This latter effect reduces the need for an \textit{ad hoc}
photoevaporation in order to get ages in agreement with astronomical observations.
\REVfirst{Unfortunately, since the dust particles have migration velocities that depend on their size, a trail of small grains
should be left along the line of sight. The smallest aggregates show an important contribution to the gas--dust opacity.
This leads to a strong extinction which could eliminate the photophoretic effect. This point is one of our most important results and,
models involving a consistent treatment of the opacity as a function of dust content, are highly desirable. Noticeably, external optically thin
disk regions could provide an environment favorable to stellar photons driven transport processes. In such a case, the net quantity of dust
delivered to the comets formation zone could not ba large enough to explain the observations, because of the vertical stratification
of grain distribution.}\\
  Besides this, using gas temperature distribution provided by our model, we confirm that amorphous grains cannot be annealed into
regions beyond $\sim 1-2$ AU, due to arguments based on phase transition kinetics at gas temperature. More importantly, we have identified the heat conductivity of
aggregates as a crucial parameter. Using realistic estimations and published experimental results for the thermal conductivity, we have
shown that the photophoretic strength can be considerably reduced and could yield to situations in which photophoretic migration
through the mid-plane could be marginal even if the disk is irradiated by a bright PMS star.\\

   Laboratory experimentations conducted by \cite{vanEymeren_wurm_2012} on ice aggregates, trapped in a cell under the combined effects
of photophoresis and thermophoresis, show that rotation induced by photophoretic
forces does not change the strength of photophoresis force, but ignores the influence of a turbulent flow.
   \cite{krauss_etal_2007} and \cite{moudens_etal_2011} have discussed, respectively the possible effects of the turbulence and of the particles
rotation. However, they considered turbulence and rotation as independent processes. \cite{krauss_etal_2007}, who discussed turbulence as a 
factor affecting the mean radial motion of dusts, concluded that turbulence essentially does not prevent the outward migration.
\cite{moudens_etal_2011} have looked to the influence of rotation of particles on themselves, induced for instance by particle to
particle collisions. We have to keep in mind that dusts are embedded in a gaseous environment, to which they are strongly coupled
(at least in the mid-plane). Following a simple picture, a turbulent flow is made of a cascade of eddies, with a distribution of sizes and
lifetimes. In such a flow, aggregates can meet eddies with rotational motion in a plane that contains the radial direction.
 In this configuration, if the thermal relaxation timescale of the particle is larger than the 
overturning timescale of the turbulent eddy, the temperature distribution at the surface of the particle could be uniformized, leading
to the removal of any photophoretic force. The global effect on dust migration through the disk depends on the intrinsic properties
of dust (thermal properties and aerodynamics drag) and on statistical properties of the involved turbulent flow. The net result
will be a convolution of the statistical distribution of grains properties (bulk material nature, porosity, size, etc) and of the 
properties of the cascade of turbulent eddies.\\
  In a first approach, the influence of the turbulence on dust trajectories could also be investigated using a particle-tracking technics as already 
employed in other contexts by \cite{supulver_lin_2000} or \cite{ciesla_sandford_2012}. Thanks to this method, applied in a 2D or 
better a 3D geometry, we could follow the trajectories of particles within the disk and even those that could be launched upwards,
pushed outward by the radiation pressure and possibly fall back onto the Nebula at different locations. A similar scenario
has been already studied by \cite{wurm_haack_2009} in their investigations concerning the outward transport of CAIs during 
FU-Orionis events. The vertical transport of grains, caused by convection and thermophoresis could be also included together with the existence of 
a quiet ``dead zone'' in which the turbulent activity should be very low.
%
%
\appendix
\section{\label{append}Description of the disk model}

   Our model of accretion disk is basically based on a generalized version of the procedure originally published by
\cite{papaloizou_terquem_1999}. More specifically, this is a 1+1D model for which the turbulence is treated in the frame
of the well known $\alpha$ formalism \citep[][]{shakura_Sunayev_1973}. During its temporal evolution, the disk is irradiated by
the star which also evolves along its Pre-Main Sequence evolutionary track.
   Both static model and disk temporal evolution programs have been implemented from scratch in \verb+FORTRAN 2008+, and 
parallelized using Open Multi-Processing (OpenMP\footnote{\url{http://openmp.org}.}). We have nicknamed 
the whole package \texttt{\textit{E}v\textit{AD}}\footnote{\textit{Ev}olutionary \textit{A}ccretion \textit{D}isk.}. 
The following sections provide a detailled description of our model.

\subsection{\label{vertstruct}Vertical structure}

    The vertical structure of the disk is governed by the equations \cite[see][]{papaloizou_terquem_1999,hure_2000} (already labeled
PT99), \cite{franck_etal_1992}
\begin{equation}
 \frac{1}{\rho} \frac{\disp\partial P}{\disp\partial z} = -\Omega^{2} z
\end{equation}
\begin{equation}
\frac{\disp\partial F}{\disp\partial z} = \frac{9}{4} \, \rho \nu \Omega^{2}
\end{equation}
\begin{equation}
\frac{\disp\partial T}{\disp\partial z} =  -\frac{\disp 3\kappa \rho}{\disp 16\sigma T^{3}} F
\end{equation}
  where $P$, $F$ and $T$ are respectively the pressure, the vertical radiative flux and the temperature; $z$ represents the
altitude above the mid-plane and $\Omega$ is the keplerian angular velocity. The density is denoted $\rho$ while $\kappa$ is the opacity of the disk's material taken in
\cite{bell_lin_1994}.
   Following PT99 the boundary condition at the external disk surface are given by 
\begin{equation}
   F_{s} = \frac{3}{8\pi} \, \dot{M}_{\rm st} \, \Omega^{2}
\end{equation}
 where $\dot{M}_{\rm st} = 3\pi <\nu> \Sigma$ with $\Sigma = \int_{-H}^{+H} \rho \, \mathrm{d}z$ (the disk
surface mass density) and $<\nu> = \int_{-H}^{+H} \rho \, \nu \, \mathrm{d}z / \Sigma$ the vertically averaged
viscosity.
  The pressure at external surface is 
\begin{equation}
   P_{s} = \frac{\disp\Omega^{2} H \tau_{ab}}{\disp\kappa_{s}}
\end{equation}
     where $H$ is the disk mid-height, $\tau_{ab}$ is the optical depth above the disk (following PT99, we have taken
$\tau_{ab}= 10^{-2}$) and $\kappa_{s}$ is the opacity at the
external surface.
  The temperature $T_{s}$ can be obtained by solving for given values of $\alpha$ and $\dot{M}_{\rm st}$ 
\begin{equation}
   2\sigma \, (T_{s}^{4}-T_{birr}^{4}) - \frac{9 \alpha k T_{s}}{8 \mu m_{H} \kappa_{s}} - \frac{3}{8\pi} \dot{M}_{\rm st} \Omega^{2} = 0
\end{equation}
 where $T_{birr}$ is given by
\begin{equation}
   T_{birr}^{4} = T_{b}^{4} + T_{irr}^{4}
\end{equation}
where $T_{irr}$ is given by Eq. (\ref{Tirr}), $T_{b}$ is the background temperature, \textit{i.e.} the temperature of the
medium in which the disk is immersed. We have chosen $T_{b}= 10$ K.
  The mid-height of the disk being not known a priori, we have to solve a Two Boundary Value Problems (TBVP). While some authors
work with relaxation algorithms \citep[][]{cannizzo_1992,milsom_etal_1994,dalessio_etal_1998,hameury_etal_1998}; we have
prefered an algorithm based on shooting methods 
\citep[][]{lin_papaloizou_1980,meyer_meyerHofmeister_1982,mineshige_osaki_1983,smak_1984,mineshige_etal_1990,rozanska_etal_1999,papaloizou_terquem_1999}
turning our TBVP into
an Initial Values Problems (IVP): the equations of the vertical structure are integrated from the surface of the disk to the mid-plan.
The 5th-order Runge-Kutta method with adaptive step length described in \cite{Num_Recipes}, already used in previous other works
\citep[see][]{papaloizou_terquem_1999,alibert_etal_2005,dodson-robinson_etal_2009}, has been employed to perform the vertical
integration. To reduce the effects of the stiffness of the set of equations, we have opted for --after several tests-- the set of
variables $u=\mathrm{ln}(\rho)$, $v= F/\sigma/T_{s}^{4}$ and $w= \mathrm{ln}(T)$ \citep[also adopted by][]{hure_2000}. 
The equation $F(z=0)= 0$ is then solved by a root finding method. Finally, the vertically averaged turbulent viscosity
$<\nu>=\int_{-H}^{+H}\rho \nu \mathrm{d}z/\Sigma$ is tabulated as a function of disk surface density 
$\Sigma= \int_{-H}^{+H}\rho \mathrm{d}z$, the steady state accretion rate $\dot{M}_{\rm st}= 3\pi <\nu> \Sigma$, the turbulent
viscosity parameter $\alpha$ and the age of the system $t$ (\textit{i.e.} providing $T_{eff}(t)$ and $R_{\star}(t)$ which are directly 
involved in the calculation. During the integration of the Eq. (\ref{Eq_evol}), these pre-built tables are interpolated ``on-the-fly'' by a
dedicated routine involving advanced B-splines technics \citep[][]{deboor_1985}.
%
\subsection{\label{secevol}Secular evolution}

  The temporal evolution of the disk is governed by Eq. (\ref{Eq_evol}), combined with boundary conditions similar
to those used by \cite{alibert_etal_2005}, which is a non-linear equation ($<\nu>$ depends on the
solution $\Sigma$). Unfortunately, there is no standard numerical method for such equation. Although explicit finite difference method is often used
\citep[\textit{e.g.}][]{papaloizo_etal_1983,ruden_lin_1986,nakamoto_nakagawa_1994,jin_sui_2010} the time-step must meet the
Courant-Friedrichs-Lewy condition \citep[see][]{courant_etal_1928} that limits the value of the time-step to small values.
Instead of an explicit finite difference scheme, we have chosen a fully implicit scheme which has the great advantage of being
unconditionally stable at least in the case of linear equations.
The obtained set of non-linear finite difference equations is solved using a multidimensional Newton--Raphson 
algorithm \footnote{\url{http://en.wikipedia.org/wiki/Newton\%27s_method}.}. Our approach
is similar to the methods used by \cite{eggleton_1971} and \cite{hameury_etal_1998}, respectively in the contexts of stellar structure or
accretion disc outburst.
  We used a decentralized finite difference formula
\begin{equation}
\left.\frac{\disp\partial f}{\disp\partial r}\right|_{i} \simeq \left.\frac{\disp\partial\xi}{\disp\partial r}\right|_{i} \,
    \frac{\disp f_{i+1} - f_{i-1}}{\disp 2\Delta\xi}
\end{equation}
  where $\xi$ is a chosen function of $r$. The initial distribution of matter being of the form $\Sigma_{0} \propto r^{-3/2}$; we have
found convenient to adopt a function like $\xi(r)= r^{p}$ where $p$ is an adjustable real parameter that tunes the distribution of points through
the disk. Finally, the evolution equation has been rewritten using $f= \Sigma <\nu> \sqrt{r}$ and $u= \sqrt{r} \,
\partial f/\partial r$.



\begin{thebibliography}{121}
\expandafter\ifx\csname natexlab\endcsname\relax\def\natexlab#1{#1}\fi
\expandafter\ifx\csname url\endcsname\relax
  \def\url#1{\texttt{#1}}\fi
\expandafter\ifx\csname urlprefix\endcsname\relax\def\urlprefix{URL }\fi

\bibitem[{{Alibert} et~al.(2005){Alibert}, {Mordasini}, {Benz}, and
  {Winisdoerffer}}]{alibert_etal_2005}
{Alibert}, Y., {Mordasini}, C., {Benz}, W., {Winisdoerffer}, C., Apr. 2005.
  {Models of giant planet formation with migration and disc evolution}. A\&A
  434, 343--353.

\bibitem[{{Allen} et~al.(2001){Allen}, {Bernstein}, and
  {Malhotra}}]{allen_etal_2001}
{Allen}, R.~L., {Bernstein}, G.~M., {Malhotra}, R., Mar. 2001. {The Edge of the
  Solar System}. ApJL 549, L241--L244.

\bibitem[{{Apai} and {Lauretta}(2010)}]{apai_lauretta_2010}
{Apai}, A., {Lauretta}, D.~S. (Eds.), 2010. {Protoplanetary Dust --
  Astrochemical and Cosmochemical Perspective}, 2nd Edition. Cambridge
  University Press.

\bibitem[{{Asplund} et~al.(2005){Asplund}, {Grevesse}, and
  {Sauval}}]{asplund_etal_2005}
{Asplund}, M., {Grevesse}, N., {Sauval}, A.~J., Sep. 2005. {The Solar Chemical
  Composition}. In: {Barnes}, III, T.~G., {Bash}, F.~N. (Eds.), Cosmic
  Abundances as Records of Stellar Evolution and Nucleosynthesis. Vol. 336 of
  Astronomical Society of the Pacific Conference Series. p.~25.

\bibitem[{{Balbus} and {Hawley}(1998)}]{balbus_hawley_1998}
{Balbus}, S.~A., {Hawley}, J.~F., Jan. 1998. {Instability, turbulence, and
  enhanced transport in accretion disks}. Reviews of Modern Physics 70, 1--53.

\bibitem[{{Bell} and {Lin}(1994)}]{bell_lin_1994}
{Bell}, K.~R., {Lin}, D.~N.~C., Jun. 1994. {Using FU Orionis outbursts to
  constrain self-regulated protostellar disk models}. ApJ 427, 987--1004.

\bibitem[{{Beresnev} et~al.(1993){Beresnev}, {Chernyak}, and
  {Fomyagin}}]{beresnev_etal_1993}
{Beresnev}, S., {Chernyak}, V., {Fomyagin}, G., 1993. Physics of Fluids 5,
  2043.

\bibitem[{{Besla} and {Wu}(2007)}]{besla_wu_2007}
{Besla}, G., {Wu}, Y., Jan. 2007. {Formation of Narrow Dust Rings in
  Circumstellar Debris Disks}. ApJ 655, 528--540.

\bibitem[{{Birnstiel}(2011)}]{birnstiel_2011}
{Birnstiel}, T., 2011. {The Evolution of Gas and Dust in Protoplanetary
  Accretion Disks}. Ph.D. thesis, PhD Thesis, 2011.

\bibitem[{{Blum} et~al.(2014){Blum}, {Gundlach}, {M{\"u}hle}, and
  {Trigo-Rodriguez}}]{blum_etal_2014}
{Blum}, J., {Gundlach}, B., {M{\"u}hle}, S., {Trigo-Rodriguez}, J.~M., Jun.
  2014. {Comets formed in solar-nebula instabilities! - An experimental and
  modeling attempt to relate the activity of comets to their formation process}
  235, 156--169.

\bibitem[{{Blum} and {Schr{\"a}pler}(2004)}]{blum_schrapler_2004}
{Blum}, J., {Schr{\"a}pler}, R., Sep. 2004. {Structure and Mechanical
  Properties of High-Porosity Macroscopic Agglomerates Formed by Random
  Ballistic Deposition}. Physical Review Letters 93~(11), 115503.

\bibitem[{{Blum} et~al.(2000){Blum}, {Wurm}, {Kempf}, {Poppe}, {Klahr},
  {Kozasa}, {Rott}, {Henning}, {Dorschner}, {Schr{\"a}pler}, {Keller},
  {Markiewicz}, {Mann}, {Gustafson}, {Giovane}, {Neuhaus}, {Fechtig},
  {Gr{\"u}n}, {Feuerbacher}, {Kochan}, {Ratke}, {El Goresy}, {Morfill},
  {Weidenschilling}, {Schwehm}, {Metzler}, and {Ip}}]{blum_etal_2000}
{Blum}, J., {Wurm}, G., {Kempf}, S., {Poppe}, T., {Klahr}, H., {Kozasa}, T.,
  {Rott}, M., {Henning}, T., {Dorschner}, J., {Schr{\"a}pler}, R., {Keller},
  H.~U., {Markiewicz}, W.~J., {Mann}, I., {Gustafson}, B.~A., {Giovane}, F.,
  {Neuhaus}, D., {Fechtig}, H., {Gr{\"u}n}, E., {Feuerbacher}, B., {Kochan},
  H., {Ratke}, L., {El Goresy}, A., {Morfill}, G., {Weidenschilling}, S.~J.,
  {Schwehm}, G., {Metzler}, K., {Ip}, W.-H., Sep. 2000. {Growth and Form of
  Planetary Seedlings: Results from a Microgravity Aggregation Experiment}.
  Physical Review Letters 85, 2426.

\bibitem[{{Bockel{\'e}e-Morvan} et~al.(2002){Bockel{\'e}e-Morvan}, {Gautier},
  {Hersant}, {Hur{\'e}}, and {Robert}}]{bockeleeMorvan_etal_2002}
{Bockel{\'e}e-Morvan}, D., {Gautier}, D., {Hersant}, F., {Hur{\'e}}, J.-M.,
  {Robert}, F., Mar. 2002. {Turbulent radial mixing in the solar nebula as the
  source of crystalline silicates in comets.} 384, 1107--1118.

\bibitem[{{Boss}(2008)}]{boss_2008}
{Boss}, A.~P., Apr. 2008. {Mixing in the solar nebula: Implications for
  isotopic heterogeneity and large-scale transport of refractory grains}. Earth
  and Planetary Science Letters 268, 102--109.

\bibitem[{{Bouvier} et~al.(2007){Bouvier}, {Alencar}, {Harries}, {Johns-Krull},
  and {Romanova}}]{bouvier_etal_2007}
{Bouvier}, J., {Alencar}, S.~H.~P., {Harries}, T.~J., {Johns-Krull}, C.~M.,
  {Romanova}, M.~M., 2007. {Magnetospheric Accretion in Classical T Tauri
  Stars}. Protostars and Planets V, 479--494.

\bibitem[{{Brownlee} et~al.(2006){Brownlee}, {Tsou}, {Al{\'e}on}, {Alexander},
  {Araki}, {Bajt}, {Baratta}, {Bastien}, {Bland}, {Bleuet}, {Borg}, {Bradley},
  {Brearley}, {Brenker}, {Brennan}, {Bridges}, {Browning}, {Brucato},
  {Bullock}, {Burchell}, {Busemann}, {Butterworth}, {Chaussidon}, {Cheuvront},
  {Chi}, {Cintala}, {Clark}, {Clemett}, {Cody}, {Colangeli}, {Cooper},
  {Cordier}, {Daghlian}, {Dai}, {D'Hendecourt}, {Djouadi}, {Dominguez},
  {Duxbury}, {Dworkin}, {Ebel}, {Economou}, {Fakra}, {Fairey}, {Fallon},
  {Ferrini}, {Ferroir}, {Fleckenstein}, {Floss}, {Flynn}, {Franchi}, {Fries},
  {Gainsforth}, {Gallien}, {Genge}, {Gilles}, {Gillet}, {Gilmour}, {Glavin},
  {Gounelle}, {Grady}, {Graham}, {Grant}, {Green}, {Grossemy}, {Grossman},
  {Grossman}, {Guan}, {Hagiya}, {Harvey}, {Heck}, {Herzog}, {Hoppe},
  {H{\"o}rz}, {Huth}, {Hutcheon}, {Ignatyev}, {Ishii}, {Ito}, {Jacob},
  {Jacobsen}, {Jacobsen}, {Jones}, {Joswiak}, {Jurewicz}, {Kearsley}, {Keller},
  {Khodja}, {Kilcoyne}, {Kissel}, {Krot}, {Langenhorst}, {Lanzirotti}, {Le},
  {Leshin}, {Leitner}, {Lemelle}, {Leroux}, {Liu}, {Luening}, {Lyon},
  {MacPherson}, {Marcus}, {Marhas}, {Marty}, {Matrajt}, {McKeegan}, {Meibom},
  {Mennella}, {Messenger}, {Messenger}, {Mikouchi}, {Mostefaoui}, {Nakamura},
  {Nakano}, {Newville}, {Nittler}, {Ohnishi}, {Ohsumi}, {Okudaira},
  {Papanastassiou}, {Palma}, {Palumbo}, {Pepin}, {Perkins}, {Perronnet},
  {Pianetta}, {Rao}, {Rietmeijer}, {Robert}, {Rost}, {Rotundi}, {Ryan},
  {Sandford}, {Schwandt}, {See}, {Schlutter}, {Sheffield-Parker},
  {Simionovici}, {Simon}, {Sitnitsky}, {Snead}, {Spencer}, {Stadermann},
  {Steele}, {Stephan}, {Stroud}, {Susini}, {Sutton}, {Suzuki}, {Taheri},
  {Taylor}, {Teslich}, {Tomeoka}, {Tomioka}, {Toppani},
  {Trigo-Rodr{\'{\i}}guez}, {Troadec}, {Tsuchiyama}, {Tuzzolino}, {Tyliszczak},
  {Uesugi}, {Velbel}, {Vellenga}, {Vicenzi}, {Vincze}, {Warren}, {Weber},
  {Weisberg}, {Westphal}, {Wirick}, {Wooden}, {Wopenka}, {Wozniakiewicz},
  {Wright}, {Yabuta}, {Yano}, {Young}, {Zare}, {Zega}, {Ziegler}, {Zimmerman},
  {Zinner}, and {Zolensky}}]{brownlee_etal_2006}
{Brownlee}, D., {Tsou}, P., {Al{\'e}on}, J., {Alexander}, C.~M.~O.~., {Araki},
  T., {Bajt}, S., {Baratta}, G.~A., {Bastien}, R., {Bland}, P., {Bleuet}, P.,
  {Borg}, J., {Bradley}, J.~P., {Brearley}, A., {Brenker}, F., {Brennan}, S.,
  {Bridges}, J.~C., {Browning}, N.~D., {Brucato}, J.~R., {Bullock}, E.,
  {Burchell}, M.~J., {Busemann}, H., {Butterworth}, A., {Chaussidon}, M.,
  {Cheuvront}, A., {Chi}, M., {Cintala}, M.~J., {Clark}, B.~C., {Clemett},
  S.~J., {Cody}, G., {Colangeli}, L., {Cooper}, G., {Cordier}, P., {Daghlian},
  C., {Dai}, Z., {D'Hendecourt}, L., {Djouadi}, Z., {Dominguez}, G., {Duxbury},
  T., {Dworkin}, J.~P., {Ebel}, D.~S., {Economou}, T.~E., {Fakra}, S.,
  {Fairey}, S.~A.~J., {Fallon}, S., {Ferrini}, G., {Ferroir}, T.,
  {Fleckenstein}, H., {Floss}, C., {Flynn}, G., {Franchi}, I.~A., {Fries}, M.,
  {Gainsforth}, Z., {Gallien}, J.-P., {Genge}, M., {Gilles}, M.~K., {Gillet},
  P., {Gilmour}, J., {Glavin}, D.~P., {Gounelle}, M., {Grady}, M.~M., {Graham},
  G.~A., {Grant}, P.~G., {Green}, S.~F., {Grossemy}, F., {Grossman}, L.,
  {Grossman}, J.~N., {Guan}, Y., {Hagiya}, K., {Harvey}, R., {Heck}, P.,
  {Herzog}, G.~F., {Hoppe}, P., {H{\"o}rz}, F., {Huth}, J., {Hutcheon}, I.~D.,
  {Ignatyev}, K., {Ishii}, H., {Ito}, M., {Jacob}, D., {Jacobsen}, C.,
  {Jacobsen}, S., {Jones}, S., {Joswiak}, D., {Jurewicz}, A., {Kearsley},
  A.~T., {Keller}, L.~P., {Khodja}, H., {Kilcoyne}, A.~L.~D., {Kissel}, J.,
  {Krot}, A., {Langenhorst}, F., {Lanzirotti}, A., {Le}, L., {Leshin}, L.~A.,
  {Leitner}, J., {Lemelle}, L., {Leroux}, H., {Liu}, M.-C., {Luening}, K.,
  {Lyon}, I., {MacPherson}, G., {Marcus}, M.~A., {Marhas}, K., {Marty}, B.,
  {Matrajt}, G., {McKeegan}, K., {Meibom}, A., {Mennella}, V., {Messenger}, K.,
  {Messenger}, S., {Mikouchi}, T., {Mostefaoui}, S., {Nakamura}, T., {Nakano},
  T., {Newville}, M., {Nittler}, L.~R., {Ohnishi}, I., {Ohsumi}, K.,
  {Okudaira}, K., {Papanastassiou}, D.~A., {Palma}, R., {Palumbo}, M.~E.,
  {Pepin}, R.~O., {Perkins}, D., {Perronnet}, M., {Pianetta}, P., {Rao}, W.,
  {Rietmeijer}, F.~J.~M., {Robert}, F., {Rost}, D., {Rotundi}, A., {Ryan}, R.,
  {Sandford}, S.~A., {Schwandt}, C.~S., {See}, T.~H., {Schlutter}, D.,
  {Sheffield-Parker}, J., {Simionovici}, A., {Simon}, S., {Sitnitsky}, I.,
  {Snead}, C.~J., {Spencer}, M.~K., {Stadermann}, F.~J., {Steele}, A.,
  {Stephan}, T., {Stroud}, R., {Susini}, J., {Sutton}, S.~R., {Suzuki}, Y.,
  {Taheri}, M., {Taylor}, S., {Teslich}, N., {Tomeoka}, K., {Tomioka}, N.,
  {Toppani}, A., {Trigo-Rodr{\'{\i}}guez}, J.~M., {Troadec}, D., {Tsuchiyama},
  A., {Tuzzolino}, A.~J., {Tyliszczak}, T., {Uesugi}, K., {Velbel}, M.,
  {Vellenga}, J., {Vicenzi}, E., {Vincze}, L., {Warren}, J., {Weber}, I.,
  {Weisberg}, M., {Westphal}, A.~J., {Wirick}, S., {Wooden}, D., {Wopenka}, B.,
  {Wozniakiewicz}, P., {Wright}, I., {Yabuta}, H., {Yano}, H., {Young}, E.~D.,
  {Zare}, R.~N., {Zega}, T., {Ziegler}, K., {Zimmerman}, L., {Zinner}, E.,
  {Zolensky}, M., Dec. 2006. {Comet 81P/Wild 2 Under a Microscope}. Science
  314, 1711--.

\bibitem[{{Brownlee}(1978)}]{brownlee_1978}
{Brownlee}, D.~E., 1978. {Microparticle studies by sampling techniques}. In:
  McDonnell, J. (Ed.), Cosmic Dust. John Wiley \& Sons, pp. 295--336.

\bibitem[{{Campins} and {Ryan}(1989)}]{campins_ryan_1989}
{Campins}, H., {Ryan}, E.~V., Jun. 1989. {The identification of crystalline
  olivine in cometary silicates}. ApJ 341, 1059--1066.

\bibitem[{{Cannizzo}(1992)}]{cannizzo_1992}
{Cannizzo}, J.~K., Jan. 1992. {Accretion disks in active galactic nuclei -
  Vertically explicit models} 385, 94--107.

\bibitem[{{Ciesla}(2007)}]{ciesla_2007}
{Ciesla}, F.~J., Oct. 2007. {Outward Transport of High-Temperature Materials
  Around the Midplane of the Solar Nebula}. Science 318, 613--.

\bibitem[{{Ciesla} and {Sandford}(2012)}]{ciesla_sandford_2012}
{Ciesla}, F.~J., {Sandford}, S.~A., Apr. 2012. {Organic Synthesis via
  Irradiation and Warming of Ice Grains in the Solar Nebula}. Science 336,
  452--.

\bibitem[{{Courant} et~al.(1928){Courant}, {Friedrichs}, and
  {Lewy}}]{courant_etal_1928}
{Courant}, R., {Friedrichs}, K., {Lewy}, H., 1928. {Über die partiellen
  Differenzengleichungen der mathematischen Physik}. Mathematische Annalen 100,
  32--74.

\bibitem[{{Crida}(2009)}]{crida_2009}
{Crida}, A., Jun. 2009. {Minimum Mass Solar Nebulae and Planetary Migration}.
  ApJ 698, 606--614.

\bibitem[{{Cunningham}(1910)}]{cunningham_1910}
{Cunningham}, J., 1910. Proceedings of the Royal Society of London Series A 83,
  357.

\bibitem[{{Cuzzi} et~al.(2003){Cuzzi}, {Davis}, and
  {Dobrovolskis}}]{cuzzi_etla_2003}
{Cuzzi}, J.~N., {Davis}, S.~S., {Dobrovolskis}, A.~R., Dec. 2003. {Blowing in
  the wind. II. Creation and redistribution of refractory inclusions in a
  turbulent protoplanetary nebula} 166, 385--402.

\bibitem[{{Cuzzi} et~al.(2014){Cuzzi}, {Estrada}, and
  {Davis}}]{cuzzi_etal_2014}
{Cuzzi}, J.~N., {Estrada}, P.~R., {Davis}, S.~S., Feb. 2014. {Utilitarian
  Opacity Model for Aggregate Particles in Protoplanetary Nebulae and Exoplanet
  Atmospheres}. ApJS 210, 21.

\bibitem[{{D'Alessio} et~al.(1998){D'Alessio}, {Cant{\"o}}, {Calvet}, and
  {Lizano}}]{dalessio_etal_1998}
{D'Alessio}, P., {Cant{\"o}}, J., {Calvet}, N., {Lizano}, S., Jun. 1998.
  {Accretion Disks around Young Objects. I. The Detailed Vertical Structure}.
  ApJ 500, 411--427.

\bibitem[{{D'Alessio} et~al.(2005){D'Alessio}, {Hartmann}, {Calvet},
  {Franco-Hern{\'a}ndez}, {Forrest}, {Sargent}, {Furlan}, {Uchida}, {Green},
  {Watson}, {Chen}, {Kemper}, {Sloan}, and {Najita}}]{dalessio_etal_2005}
{D'Alessio}, P., {Hartmann}, L., {Calvet}, N., {Franco-Hern{\'a}ndez}, R.,
  {Forrest}, W.~J., {Sargent}, B., {Furlan}, E., {Uchida}, K., {Green}, J.~D.,
  {Watson}, D.~M., {Chen}, C.~H., {Kemper}, F., {Sloan}, G.~C., {Najita}, J.,
  Mar. 2005. {The Truncated Disk of CoKu Tau/4} 621, 461--472.

\bibitem[{{de Boor}(1985)}]{deboor_1985}
{de Boor}, C., 1985. {A Practical Guide to Splines}, 3rd Edition. Springer.

\bibitem[{{Degl'Innocenti} et~al.(2008){Degl'Innocenti}, {Prada Moroni},
  {Marconi}, and {Ruoppo}}]{deglinnocenti_etal_2008}
{Degl'Innocenti}, S., {Prada Moroni}, P.~G., {Marconi}, M., {Ruoppo}, A., Aug.
  2008. {The FRANEC stellar evolutionary code}. ApSS 316, 25--30.

\bibitem[{{Dell'Omodarme} et~al.(2012){Dell'Omodarme}, {Valle},
  {Degl'Innocenti}, and {Prada Moroni}}]{dellomodarme_etal_2012}
{Dell'Omodarme}, M., {Valle}, G., {Degl'Innocenti}, S., {Prada Moroni}, P.~G.,
  Apr. 2012. {The Pisa Stellar Evolution Data Base for low-mass stars}. A\&A
  540, A26.

\bibitem[{{Dodson-Robinson} et~al.(2009){Dodson-Robinson}, {Willacy},
  {Bodenheimer}, {Turner}, and {Beichman}}]{dodson-robinson_etal_2009}
{Dodson-Robinson}, S.~E., {Willacy}, K., {Bodenheimer}, P., {Turner}, N.~J.,
  {Beichman}, C.~A., Apr. 2009. {Ice lines, planetesimal composition and solid
  surface density in the solar nebula} 200, 672--693.

\bibitem[{{Eggleton}(1971)}]{eggleton_1971}
{Eggleton}, P.~P., 1971. {The evolution of low mass stars}. MNRAS 151, 351.

\bibitem[{{Fam{\'a}} et~al.(2010){Fam{\'a}}, {Loeffler}, {Raut}, and
  {Baragiola}}]{fama_etal_2010}
{Fam{\'a}}, M., {Loeffler}, M.~J., {Raut}, U., {Baragiola}, R.~A., May 2010.
  {Radiation-induced amorphization of crystalline ice} 207, 314--319.

\bibitem[{{Ferreira} et~al.(2006){Ferreira}, {Dougados}, and
  {Cabrit}}]{ferreira_etal_2006}
{Ferreira}, J., {Dougados}, C., {Cabrit}, S., Jul. 2006. {Which jet launching
  mechanism(s) in T Tauri stars?} 453, 785--796.

\bibitem[{{Fouchet} et~al.(2012){Fouchet}, {Alibert}, {Mordasini}, and
  {Benz}}]{fouchet_etal_2012}
{Fouchet}, L., {Alibert}, Y., {Mordasini}, C., {Benz}, W., Apr. 2012. {Effects
  of disk irradiation on planet population synthesis}. A\&A 540, A107.

\bibitem[{{Frank} et~al.(1992){Frank}, {King}, and {Raine}}]{franck_etal_1992}
{Frank}, J., {King}, A., {Raine}, D., 1992. {Accretion power in astrophysics}.
  Cambridge Uni. Press., San Diego.

\bibitem[{{Gail}(2001)}]{gail_2001}
{Gail}, H.-P., Oct. 2001. {Radial mixing in protoplanetary accretion disks. I.
  Stationary disc models with annealing and carbon combustion} 378, 192--213.

\bibitem[{{Greenberg}(1985)}]{greenberg_1985}
{Greenberg}, J.~M., 1985. {The chemical and physical evolution of interstellar
  dust}. Physica Scripta Volume T 11, 14--26.

\bibitem[{{Gundlach} and {Blum}(2012)}]{gundlach_blum_2012}
{Gundlach}, B., {Blum}, J., Jun. 2012. {Outgassing of icy bodies in the Solar
  System - II: Heat transport in dry, porous surface dust layers} 219,
  618--629.

\bibitem[{{G{\"u}ttler} et~al.(2010){G{\"u}ttler}, {Blum}, {Zsom}, {Ormel}, and
  {Dullemond}}]{guttler_etal_2010}
{G{\"u}ttler}, C., {Blum}, J., {Zsom}, A., {Ormel}, C.~W., {Dullemond}, C.~P.,
  Apr. 2010. {The outcome of protoplanetary dust growth: pebbles, boulders, or
  planetesimals?. I. Mapping the zoo of laboratory collision experiments}. A\&A
  513, A56.

\bibitem[{{Haisch} et~al.(2001){Haisch}, {Lada}, and {Lada}}]{haisch_etal_2001}
{Haisch}, Jr., K.~E., {Lada}, E.~A., {Lada}, C.~J., Jun. 2001. {Disk
  Frequencies and Lifetimes in Young Clusters}. ApJ 553, L153--L156.

\bibitem[{{Hameury} et~al.(1998){Hameury}, {Menou}, {Dubus}, {Lasota}, and
  {Hure}}]{hameury_etal_1998}
{Hameury}, J.-M., {Menou}, K., {Dubus}, G., {Lasota}, J.-P., {Hure}, J.-M.,
  Aug. 1998. {Accretion disc outbursts: a new version of an old model}. MNRAS
  298, 1048--1060.

\bibitem[{{Harker} and {Desch}(2002)}]{harker_desch_2002}
{Harker}, D.~E., {Desch}, S.~J., Feb. 2002. {Annealing of Silicate Dust by
  Nebular Shocks at 10 AU}. ApJL 565, L109--L112.

\bibitem[{{Hayashi}(1981)}]{hayashi_1981}
{Hayashi}, C., 1981. {Structure of the Solar Nebula, Growth and Decay of
  Magnetic Fields and Effects of Magnetic and Turbulent Viscosities on the
  Nebula}. Progress of Theoretical Physics Supplement 70, 35--53.

\bibitem[{{Horai} and {Simmons}(1969)}]{horai_simmons_1969}
{Horai}, K., {Simmons}, G., 1969. Earth and Planetary Science Letters 6,
  359--368.

\bibitem[{{Hueso} and {Guillot}(2005)}]{hueso_guillot_2005}
{Hueso}, R., {Guillot}, T., Nov. 2005. {Evolution of protoplanetary disks:
  constraints from DM Tauri and GM Aurigae}. A\&A 442, 703--725.

\bibitem[{{Hughes} et~al.(2010){Hughes}, {Andrews}, {Wilner}, {Meyer},
  {Carpenter}, {Qi}, {Hales}, {Casassus}, {Hogerheijde}, {Mamajek}, {Wolf},
  {Henning}, and {Silverstone}}]{hughes_etal_2010}
{Hughes}, A.~M., {Andrews}, S.~M., {Wilner}, D.~J., {Meyer}, M.~R.,
  {Carpenter}, J.~M., {Qi}, C., {Hales}, A.~S., {Casassus}, S., {Hogerheijde},
  M.~R., {Mamajek}, E.~E., {Wolf}, S., {Henning}, T., {Silverstone}, M.~D.,
  Sep. 2010. {Structure and Composition of Two Transitional Circumstellar Disks
  in Corona Australis}. AJ 140, 887--896.

\bibitem[{{Hur{\'e}}(2000)}]{hure_2000}
{Hur{\'e}}, J.-M., Jun. 2000. {On the transition to self-gravity in low mass
  AGN and YSO accretion discs}. A\&A 358, 378--394.

\bibitem[{{Hutchins} et~al.(1995){Hutchins}, {Harper}, and
  {Felder}}]{hutchins_etal_1995}
{Hutchins}, D.~K., {Harper}, M.~H., {Felder}, R.~L., 1995. Aerosol Science and
  Technology 22, 202.

\bibitem[{{Jin} and {Sui}(2010)}]{jin_sui_2010}
{Jin}, L., {Sui}, N., Feb. 2010. {The Evolution of the Solar Nebula I.
  Evolution of the Global Properties and Planet Masses}. ApJ 710, 1179--1194.

\bibitem[{{Keller} and {Gail}(2004)}]{keller_gail_2004}
{Keller}, C., {Gail}, H.-P., Mar. 2004. {Radial mixing in protoplanetary
  accretion disks. VI. Mixing by large-scale radial flows}. A\&A 415,
  1177--1185.

\bibitem[{{Kelley} and {Wooden}(2009)}]{kelley_wooden_2009}
{Kelley}, M.~S., {Wooden}, D.~H., Aug. 2009. {The composition of dust in
  Jupiter-family comets inferred from infrared spectroscopy} 57, 1133--1145.

\bibitem[{{Kemper} et~al.(2004){Kemper}, {Vriend}, and
  {Tielens}}]{kemper_etal_2004}
{Kemper}, F., {Vriend}, W.~J., {Tielens}, A.~G.~G.~M., Jul. 2004. {The Absence
  of Crystalline Silicates in the Diffuse Interstellar Medium}. ApJ 609,
  826--837.

\bibitem[{{Kogel} et~al.(2006){Kogel}, {Trivedi}, {Barker}, and
  {Krukowski}}]{kogel_etal_2006}
{Kogel}, J.~E., {Trivedi}, N.~C., {Barker}, J.~M., {Krukowski}, S., 2006.
  {Industrial Minerals \& Rocks (Book-CD Set), 7th edition}, 7th Edition.
  Society for Mining, Metallurgy, and Exploration, Englewood, CO 80112.

\bibitem[{{Krause} et~al.(2011){Krause}, {Blum}, {Skorov}, and
  {Trieloff}}]{krause_etal_2011}
{Krause}, M., {Blum}, J., {Skorov}, Y.~V., {Trieloff}, M., Jul. 2011. {Thermal
  conductivity measurements of porous dust aggregates: I. Technique, model and
  first results} 214, 286--296.

\bibitem[{{Krauss} and {Wurm}(2005)}]{krauss_wurm_2005}
{Krauss}, O., {Wurm}, G., Sep. 2005. {Photophoresis and the Pile-up of Dust in
  Young Circumstellar Disks}. ApJ 630, 1088--1092.

\bibitem[{{Krauss} et~al.(2007){Krauss}, {Wurm}, {Mousis}, {Petit}, {Horner},
  and {Alibert}}]{krauss_etal_2007}
{Krauss}, O., {Wurm}, G., {Mousis}, O., {Petit}, J.-M., {Horner}, J.,
  {Alibert}, Y., Feb. 2007. {The photophoretic sweeping of dust in transient
  protoplanetary disks}. A\&A 462, 977--987.

\bibitem[{{Le Guillou} et~al.(2013){Le Guillou}, {Remusat}, {Bernard},
  {Brearley}, and {Leroux}}]{leguillou_etal_2013}
{Le Guillou}, C., {Remusat}, L., {Bernard}, S., {Brearley}, A.~J., {Leroux},
  H., Sep. 2013. {Amorphization and D/H fractionation of kerogens during
  experimental electron irradiation: Comparison with chondritic organic matter}
  226, 101--110.

\bibitem[{{Lenzuni} et~al.(1995){Lenzuni}, {Gail}, and
  {Henning}}]{lenzuni_etal_1995}
{Lenzuni}, P., {Gail}, H.-P., {Henning}, T., Jul. 1995. {Dust Evaporation in
  Protostellar Cores}. ApJ 447, 848.

\bibitem[{{Lide} D.~R.(2005)}]{Handbook_2005}
{Lide} D.~R., e., 2005. Handbook of Chemistry and Physics, 1st Edition.

\bibitem[{{Lin} and {Papaloizou}(1980)}]{lin_papaloizou_1980}
{Lin}, D.~N.~C., {Papaloizou}, J., Apr. 1980. {On the structure and evolution
  of the primordial solar nebula} 191, 37--48.

\bibitem[{{Lindsay} et~al.(2013){Lindsay}, {Wooden}, {Harker}, {Kelley},
  {Woodward}, and {Murphy}}]{lindsay_etal_2013}
{Lindsay}, S.~S., {Wooden}, D.~H., {Harker}, D.~E., {Kelley}, M.~S.,
  {Woodward}, C.~E., {Murphy}, J.~R., Mar. 2013. {Absorption Efficiencies of
  Forsterite. I. Discrete Dipole Approximation Explorations in Grain Shape and
  Size} 766, 54.

\bibitem[{{Loesche} et~al.(2013){Loesche}, {Wurm}, {Teiser}, {Friedrich}, and
  {Bischoff}}]{loesche_etal_2013a}
{Loesche}, C., {Wurm}, G., {Teiser}, J., {Friedrich}, J.~M., {Bischoff}, A.,
  Dec. 2013. {Photophoretic Strength on Chondrules. 1. Modeling} 778, 101.

\bibitem[{{Lutro}(2012)}]{lutro_2012}
{Lutro}, H.~F., 2012. {The Effect of Thermophoresis on the Particle Deposition
  on a Cylinder}. Ph.D. thesis, NTNU -- Trondheim-- Norwegian University of
  Science and Technology.

\bibitem[{{Lynden-Bell} and {Pringle}(1974)}]{lynden-Bell_Pringle_1974}
{Lynden-Bell}, D., {Pringle}, J.~E., Sep. 1974. {The evolution of viscous discs
  and the origin of the nebular variables.} MNRAS 168, 603--637.

\bibitem[{{Meakin} and {Donn}(1988)}]{meakin_donn_1988}
{Meakin}, P., {Donn}, B., Jun. 1988. {Aerodynamic properties of fractal grains
  - Implications for the primordial solar nebula} 329, L39--L41.

\bibitem[{{Meyer} and {Meyer-Hofmeister}(1982)}]{meyer_meyerHofmeister_1982}
{Meyer}, F., {Meyer-Hofmeister}, E., Feb. 1982. {Vertical structure of
  accretion disks} 106, 34--42.

\bibitem[{{Milsom} et~al.(1994){Milsom}, {Chen}, and {Taam}}]{milsom_etal_1994}
{Milsom}, J.~A., {Chen}, X., {Taam}, R.~E., Feb. 1994. {The vertical structure
  and stability of accretion disks surrounding black holes and neutron stars}
  421, 668--676.

\bibitem[{{Mineshige} and {Osaki}(1983)}]{mineshige_osaki_1983}
{Mineshige}, S., {Osaki}, Y., 1983. {Disk-instability model for outbursts of
  dwarf novae Time-dependent formulation and one-zone model} 35, 377--396.

\bibitem[{{Mineshige} et~al.(1990){Mineshige}, {Tuchman}, and
  {Wheeler}}]{mineshige_etal_1990}
{Mineshige}, S., {Tuchman}, Y., {Wheeler}, J.~C., Aug. 1990. {Structure and
  Evolution of Irradiated Accretion Disks. II. Dynamical Evolution of a
  Thermally Unstable Torus} 359, 176.

\bibitem[{{Moudens} et~al.(2011){Moudens}, {Mousis}, {Petit}, {Wurm},
  {Cordier}, and {Charnoz}}]{moudens_etal_2011}
{Moudens}, A., {Mousis}, O., {Petit}, J.-M., {Wurm}, G., {Cordier}, D.,
  {Charnoz}, S., Jul. 2011. {Photophoretic transport of hot minerals in the
  solar nebula}. A\&A 531, A106.

\bibitem[{{Mousis} et~al.(2007){Mousis}, {Petit}, {Wurm}, {Krauss}, {Alibert},
  and {Horner}}]{mousis_etal_2007}
{Mousis}, O., {Petit}, J.-M., {Wurm}, G., {Krauss}, O., {Alibert}, Y.,
  {Horner}, J., May 2007. {Photophoresis as a source of hot minerals in
  comets}. A\&A 466, L9--L12.

\bibitem[{{Mumma} and {Charnley}(2011)}]{mumma_charnley_2011}
{Mumma}, M.~J., {Charnley}, S.~B., Sep. 2011. {The Chemical Composition of
  Comets-Emerging Taxonomies and Natal Heritage}. ARA\&A 49, 471--524.

\bibitem[{{Nakamoto} and {Nakagawa}(1994)}]{nakamoto_nakagawa_1994}
{Nakamoto}, T., {Nakagawa}, Y., Feb. 1994. {Formation, early evolution, and
  gravitational stability of protoplanetary disks}. ApJ 421, 640--650.

\bibitem[{{Ogliore} et~al.(2009){Ogliore}, {Westphal}, {Gainsforth},
  {Butterworth}, {Fakra}, and {Marcus}}]{ogliore_etal_2009}
{Ogliore}, R.~C., {Westphal}, A.~J., {Gainsforth}, Z., {Butterworth}, A.~L.,
  {Fakra}, S.~C., {Marcus}, M.~A., Nov. 2009. {Nebular mixing constrained by
  the Stardust samples}. Meteoritics and Planetary Science 44, 1675--1681.

\bibitem[{{Ollivier} et~al.(2009){Ollivier}, {Encrenaz}, {Roques}, {Selsis},
  and {Casoli}}]{ollivier_etal_2009}
{Ollivier}, M., {Encrenaz}, T., {Roques}, F., {Selsis}, F., {Casoli}, F., 2009.
  {Planetary Systems -- Detection, Formation and Habitability of Extrasolar
  Planets}, 3rd Edition. Springer-Verlag, Berlin, Heidelberg.

\bibitem[{{Ormel} et~al.(2007){Ormel}, {Spaans}, and
  {Tielens}}]{ormel_etal_2007}
{Ormel}, C.~W., {Spaans}, M., {Tielens}, A.~G.~G.~M., Jan. 2007. {Dust
  coagulation in protoplanetary disks: porosity matters}. A\&A 461, 215--232.

\bibitem[{{Owen}(2006)}]{owen_2006}
{Owen}, T.~C., Dec. 2006. {The Origin of Nitrogen Atmospheres on Earth and
  Titan}. AGU Fall Meeting Abstracts, A5+.

\bibitem[{{Papaloizou} et~al.(1983){Papaloizou}, {Faulkner}, and
  {Lin}}]{papaloizo_etal_1983}
{Papaloizou}, J., {Faulkner}, J., {Lin}, D.~N.~C., Nov. 1983. {On the evolution
  of accretion disc flow in cataclysmic variables. II - The existence and
  nature of the collective relaxation oscillations in dwarf nova systems}.
  MNRAS 205, 487--513.

\bibitem[{{Papaloizou} and {Terquem}(1999)}]{papaloizou_terquem_1999}
{Papaloizou}, J.~C.~B., {Terquem}, C., Aug. 1999. {Critical Protoplanetary Core
  Masses in Protoplanetary Disks and the Formation of Short-Period Giant
  Planets}. ApJ 521, 823--838.

\bibitem[{{Pascucci} and {Tachibana}(2010)}]{pascucci_tachibana_2010}
{Pascucci}, I., {Tachibana}, S., Feb. 2010. {The Clearing of Protoplanetary
  Disks and of the Protosolar Nebula}. pp. 263--298.

\bibitem[{{Pietrinferni} et~al.(2004){Pietrinferni}, {Cassisi}, {Salaris}, and
  {Castelli}}]{pietrinferni_etal_2004}
{Pietrinferni}, A., {Cassisi}, S., {Salaris}, M., {Castelli}, F., Sep. 2004. {A
  Large Stellar Evolution Database for Population Synthesis Studies. I. Scaled
  Solar Models and Isochrones}. ApJ 612, 168--190.

\bibitem[{{Pontoppidan} et~al.(2008){Pontoppidan}, {Blake}, {van Dishoeck},
  {Smette}, {Ireland}, and {Brown}}]{pontoppidan_etal_2008}
{Pontoppidan}, K.~M., {Blake}, G.~A., {van Dishoeck}, E.~F., {Smette}, A.,
  {Ireland}, M.~J., {Brown}, J., Sep. 2008. {Spectroastrometric Imaging of
  Molecular Gas within Protoplanetary Disk Gaps}. ApJ 684, 1323--1329.

\bibitem[{{Presley} and {Christensen}(1997)}]{presley_christensen_1997}
{Presley}, M.~A., {Christensen}, P.~R., Mar. 1997. {Thermal conductivity
  measurements of particulate materials 1. A review}. JGR 102, 6535--6550.

\bibitem[{{Press} et~al.(1992){Press}, {Teukolsky}, {Vetterling}, and
  {Flannery}}]{Num_Recipes}
{Press}, W., {Teukolsky}, S., {Vetterling}, W., {Flannery}, B., 1992. Numerical
  Recipes in Fortran 77. Cambridge University Press.

\bibitem[{{Pringle}(1981)}]{pringle_1981}
{Pringle}, J.~E., 1981. {Accretion discs in astrophysics}. ARA\&A 19, 137--162.

\bibitem[{{Reif}(1967)}]{reif_1967}
{Reif}, F., 1967. Berkeley Physics Course: Statistical physics. Berkeley
  Physics Course. McGraw-Hill.

\bibitem[{{R{\'o}{\.z}a{\'n}ska} et~al.(1999){R{\'o}{\.z}a{\'n}ska}, {Czerny},
  {{\.Z}ycki}, and {Pojma{\'n}ski}}]{rozanska_etal_1999}
{R{\'o}{\.z}a{\'n}ska}, A., {Czerny}, B., {{\.Z}ycki}, P.~T., {Pojma{\'n}ski},
  G., May 1999. {Vertical structure of accretion discs with hot coronae in
  active galactic nuclei} 305, 481--491.

\bibitem[{{Ruden} and {Lin}(1986)}]{ruden_lin_1986}
{Ruden}, S.~P., {Lin}, D.~N.~C., Sep. 1986. {The global evolution of the
  primordial solar nebula}. ApJ 308, 883--901.

\bibitem[{{Russell}(1935)}]{russell_1935}
{Russell}, H.~W., 1935. J. Am. Ceram. Soc. 18, 1.

\bibitem[{{Shakura} and {Sunyaev}(1973)}]{shakura_Sunayev_1973}
{Shakura}, N.~I., {Sunyaev}, R.~A., 1973. A\&A 24, 337.

\bibitem[{{Shu} et~al.(1996){Shu}, {Shang}, and {Lee}}]{shu_etal_1996}
{Shu}, F.~H., {Shang}, H., {Lee}, T., Mar. 1996. {Toward an Astrophysical
  Theory of Chondrites}. Science 271, 1545--1552.

\bibitem[{{Sicilia-Aguilar} et~al.(2006){Sicilia-Aguilar}, {Hartmann},
  {Calvet}, {Megeath}, {Muzerolle}, {Allen}, {D'Alessio}, {Mer{\'{\i}}n},
  {Stauffer}, {Young}, and {Lada}}]{siciliaaguilar_etal_2006}
{Sicilia-Aguilar}, A., {Hartmann}, L., {Calvet}, N., {Megeath}, S.~T.,
  {Muzerolle}, J., {Allen}, L., {D'Alessio}, P., {Mer{\'{\i}}n}, B.,
  {Stauffer}, J., {Young}, E., {Lada}, C., Feb. 2006. {Disk Evolution in Cep
  OB2: Results from the Spitzer Space Telescope} 638, 897--919.

\bibitem[{{Smak}(1984)}]{smak_1984}
{Smak}, J., 1984. {Accretion in cataclysmic binaries. IV - Accretion disks in
  dwarf novae} 34, 161--189.

\bibitem[{{Stahler} and {Palla}(2004)}]{stahler_palla_2004}
{Stahler}, S.~W., {Palla}, F., 2004. {The Formation of Stars}, wiley Edition.
  New York.

\bibitem[{{Supulver} and {Lin}(2000)}]{supulver_lin_2000}
{Supulver}, K.~D., {Lin}, D.~N.~C., Aug. 2000. {Formation of Icy Planetesimals
  in a Turbulent Solar Nebula} 146, 525--540.

\bibitem[{{Teiser} and {Dodson-Robinson}(2013)}]{teiser_dodsonrobinson2013}
{Teiser}, J., {Dodson-Robinson}, S.~E., Jul. 2013. {Photophoresis boosts giant
  planet formation}. A\&A 555, A98.

\bibitem[{{Thalmann} et~al.(2010){Thalmann}, {Grady}, {Goto}, {Wisniewski},
  {Janson}, {Henning}, {Fukagawa}, {Honda}, {Mulders}, {Min},
  {Moro-Mart{\'{\i}}n}, {McElwain}, {Hodapp}, {Carson}, {Abe}, {Brandner},
  {Egner}, {Feldt}, {Fukue}, {Golota}, {Guyon}, {Hashimoto}, {Hayano},
  {Hayashi}, {Hayashi}, {Ishii}, {Kandori}, {Knapp}, {Kudo}, {Kusakabe},
  {Kuzuhara}, {Matsuo}, {Miyama}, {Morino}, {Nishimura}, {Pyo}, {Serabyn},
  {Shibai}, {Suto}, {Suzuki}, {Takami}, {Takato}, {Terada}, {Tomono}, {Turner},
  {Watanabe}, {Yamada}, {Takami}, {Usuda}, and {Tamura}}]{thalmann_etal_2010}
{Thalmann}, C., {Grady}, C.~A., {Goto}, M., {Wisniewski}, J.~P., {Janson}, M.,
  {Henning}, T., {Fukagawa}, M., {Honda}, M., {Mulders}, G.~D., {Min}, M.,
  {Moro-Mart{\'{\i}}n}, A., {McElwain}, M.~W., {Hodapp}, K.~W., {Carson}, J.,
  {Abe}, L., {Brandner}, W., {Egner}, S., {Feldt}, M., {Fukue}, T., {Golota},
  T., {Guyon}, O., {Hashimoto}, J., {Hayano}, Y., {Hayashi}, M., {Hayashi}, S.,
  {Ishii}, M., {Kandori}, R., {Knapp}, G.~R., {Kudo}, T., {Kusakabe}, N.,
  {Kuzuhara}, M., {Matsuo}, T., {Miyama}, S., {Morino}, J.-I., {Nishimura}, T.,
  {Pyo}, T.-S., {Serabyn}, E., {Shibai}, H., {Suto}, H., {Suzuki}, R.,
  {Takami}, M., {Takato}, N., {Terada}, H., {Tomono}, D., {Turner}, E.~L.,
  {Watanabe}, M., {Yamada}, T., {Takami}, H., {Usuda}, T., {Tamura}, M., Aug.
  2010. {Imaging of a Transitional Disk Gap in Reflected Light: Indications of
  Planet Formation Around the Young Solar Analog LkCa 15}. ApJL 718, L87--L91.

\bibitem[{{Tielens} et~al.(2005){Tielens}, {Waters}, and
  {Bernatowicz}}]{tielens_etal_2005}
{Tielens}, A.~G.~G.~M., {Waters}, L.~B.~F.~M., {Bernatowicz}, T.~J., Dec. 2005.
  {Origin and Evolution of Dust in Circumstellar and Interstellar
  Environments}. In: {Krot}, A.~N., {Scott}, E.~R.~D., {Reipurth}, B. (Eds.),
  Chondrites and the Protoplanetary Disk. Vol. 341 of Astronomical Society of
  the Pacific Conference Series. p. 605.

\bibitem[{{Tognelli} et~al.(2012){Tognelli}, {Degl'Innocenti}, and {Prada
  Moroni}}]{tognelli_etal_2012}
{Tognelli}, E., {Degl'Innocenti}, S., {Prada Moroni}, P.~G., Dec. 2012.
  {$^{7}$Li surface abundance in pre-main sequence stars. Testing theory
  against clusters and binary systems}. A\&A 548, A41.

\bibitem[{{Tognelli} et~al.(2011){Tognelli}, {Prada Moroni}, and
  {Degl'Innocenti}}]{tognelli_etal_2011}
{Tognelli}, E., {Prada Moroni}, P.~G., {Degl'Innocenti}, S., Sep. 2011. {The
  Pisa pre-main sequence tracks and isochrones. A database covering a wide
  range of Z, Y, mass, and age values}. A\&A 533, A109.

\bibitem[{{Tsiganis} et~al.(2005){Tsiganis}, {Gomes}, {Morbidelli}, and
  {Levison}}]{tsiganis_etal_2005}
{Tsiganis}, K., {Gomes}, R., {Morbidelli}, A., {Levison}, H.~F., May 2005.
  {Origin of the orbital architecture of the giant planets of the Solar
  System}. Nature 435, 459--461.

\bibitem[{{Turner} et~al.(2012){Turner}, {Choukroun}, {Castillo-Rogez}, and
  {Bryden}}]{turner_etal_2012}
{Turner}, N.~J., {Choukroun}, M., {Castillo-Rogez}, J., {Bryden}, G., Apr.
  2012. {A Hot Gap around Jupiter's Orbit in the Solar Nebula}. ApJ 748, 92.

\bibitem[{{van Eymeren} and {Wurm}(2012)}]{vanEymeren_wurm_2012}
{van Eymeren}, J., {Wurm}, G., Feb. 2012. {The implications of particle
  rotation on the effect of photophoresis}. MNRAS 420, 183--186.

\bibitem[{{Veras} and {Armitage}(2004)}]{veras_armitage_2004}
{Veras}, D., {Armitage}, P.~J., Jan. 2004. {Outward migration of extrasolar
  planets to large orbital radii}. MNRAS 347, 613--624.

\bibitem[{{von Borstel} and {Blum}(2012)}]{vonBorstel_blum_2012}
{von Borstel}, I., {Blum}, J., Dec. 2012. {Photophoresis of dust aggregates in
  protoplanetary disks}. A\&A 548, A96.

\bibitem[{{Wehrstedt} and {Gail}(2002)}]{wehrstedt_gail_2002}
{Wehrstedt}, M., {Gail}, H.-P., Apr. 2002. {Radial mixing in protoplanetary
  accretion disks. II. Time dependent disk models with annealing and carbon
  combustion} 385, 181--204.

\bibitem[{{Weidenschilling}(1977{\natexlab{a}})}]{weidenschilling_1977}
{Weidenschilling}, S.~J., Sep. 1977{\natexlab{a}}. {The distribution of mass in
  the planetary system and solar nebula}. Astrophys. Space Sci. 51, 153--158.

\bibitem[{{Weidenschilling}(1977{\natexlab{b}})}]{weidenschilling_1977b}
{Weidenschilling}, S.~J., Sep. 1977{\natexlab{b}}. {The distribution of mass in
  the planetary system and solar nebula}. Astrophysics and Space Science 51,
  153--158.

\bibitem[{{Weidenschilling}(1997)}]{weidenschilling_1997}
{Weidenschilling}, S.~J., Jun. 1997. {The Origin of Comets in the Solar Nebula:
  A Unified Model} 127, 290--306.

\bibitem[{{Wooden} et~al.(2000){Wooden}, {Butner}, {Harker}, and
  {Woodward}}]{wooden_etal_2000}
{Wooden}, D.~H., {Butner}, H.~M., {Harker}, D.~E., {Woodward}, C.~E., Jan.
  2000. {Mg-Rich Silicate Crystals in Comet Hale-Bopp: ISM Relics or Solar
  Nebula Condensates?} Icarus 143, 126--137.

\bibitem[{{Wooden} et~al.(1999){Wooden}, {Harker}, {Woodward}, {Butner},
  {Koike}, {Witteborn}, and {McMurtry}}]{wooden_etal_1999}
{Wooden}, D.~H., {Harker}, D.~E., {Woodward}, C.~E., {Butner}, H.~M., {Koike},
  C., {Witteborn}, F.~C., {McMurtry}, C.~W., Jun. 1999. {Silicate Mineralogy of
  the Dust in the Inner Coma of Comet C/1995 01 (Hale-Bopp) Pre- and
  Postperihelion}. ApJ 517, 1034--1058.

\bibitem[{{Wurm} and {Haack}(2009)}]{wurm_haack_2009}
{Wurm}, G., {Haack}, H., Jul. 2009. {Outward transport of CAIs during
  FU-Orionis events}. Meteoritics and Planetary Science 44, 689--699.

\bibitem[{{Wurm} and {Krauss}(2006)}]{wurm_krauss_2006}
{Wurm}, G., {Krauss}, O., Feb. 2006. {Concentration and sorting of chondrules
  and CAIs in the late Solar Nebula} 180, 487--495.

\bibitem[{{Wurm} et~al.(2010){Wurm}, {Teiser}, {Bischoff}, {Haack}, and
  {Roszjar}}]{wurm_etal_2010}
{Wurm}, G., {Teiser}, J., {Bischoff}, A., {Haack}, H., {Roszjar}, J., Jul.
  2010. {Experiments on the photophoretic motion of chondrules and dust
  aggregates--Indications for the transport of matter in protoplanetary disks}
  208, 482--491.

\bibitem[{{Wurm} et~al.(2013){Wurm}, {Trieloff}, and {Rauer}}]{wurm_etal_2013}
{Wurm}, G., {Trieloff}, M., {Rauer}, H., May 2013. {Photophoretic Separation of
  Metals and Silicates: The Formation of Mercury-like Planets and Metal
  Depletion in Chondrites}. ApJ 769, 78.

\bibitem[{{Xu} et~al.(2004){Xu}, {Shankland}, {Linhardt}, {Rubie},
  {Langenhorst}, and {Klasinski}}]{xu_etal_2004}
{Xu}, Y., {Shankland}, T.~J., {Linhardt}, S., {Rubie}, D.~C., {Langenhorst},
  F., {Klasinski}, K., 2004. {Thermal diffusivity and conductivity of olivine,
  wadsleyite and ringwoodite to 20 GPa and 1373 K}. Physics of the Earth and
  Planetary Interiors 143, 321–336.

\bibitem[{{Young}(2011)}]{young_2011}
{Young}, J.~B., 2011. {Thermophoresis of a spherical particle: Reassessment,
  clarication, and new analysis}. Aerosol Science and Technology 45, 927--948.

\bibitem[{{Zolensky} et~al.(2006){Zolensky}, {Zega}, {Yano}, {Wirick},
  {Westphal}, {Weisberg}, {Weber}, {Warren}, {Velbel}, {Tsuchiyama}, {Tsou},
  {Toppani}, {Tomioka}, {Tomeoka}, {Teslich}, {Taheri}, {Susini}, {Stroud},
  {Stephan}, {Stadermann}, {Snead}, {Simon}, {Simionovici}, {See}, {Robert},
  {Rietmeijer}, {Rao}, {Perronnet}, {Papanastassiou}, {Okudaira}, {Ohsumi},
  {Ohnishi}, {Nakamura-Messenger}, {Nakamura}, {Mostefaoui}, {Mikouchi},
  {Meibom}, {Matrajt}, {Marcus}, {Leroux}, {Lemelle}, {Le}, {Lanzirotti},
  {Langenhorst}, {Krot}, {Keller}, {Kearsley}, {Joswiak}, {Jacob}, {Ishii},
  {Harvey}, {Hagiya}, {Grossman}, {Grossman}, {Graham}, {Gounelle}, {Gillet},
  {Genge}, {Flynn}, {Ferroir}, {Fallon}, {Ebel}, {Dai}, {Cordier}, {Clark},
  {Chi}, {Butterworth}, {Brownlee}, {Bridges}, {Brennan}, {Brearley},
  {Bradley}, {Bleuet}, {Bland}, and {Bastien}}]{zolensky_etla_2006}
{Zolensky}, M.~E., {Zega}, T.~J., {Yano}, H., {Wirick}, S., {Westphal}, A.~J.,
  {Weisberg}, M.~K., {Weber}, I., {Warren}, J.~L., {Velbel}, M.~A.,
  {Tsuchiyama}, A., {Tsou}, P., {Toppani}, A., {Tomioka}, N., {Tomeoka}, K.,
  {Teslich}, N., {Taheri}, M., {Susini}, J., {Stroud}, R., {Stephan}, T.,
  {Stadermann}, F.~J., {Snead}, C.~J., {Simon}, S.~B., {Simionovici}, A.,
  {See}, T.~H., {Robert}, F., {Rietmeijer}, F.~J.~M., {Rao}, W., {Perronnet},
  M.~C., {Papanastassiou}, D.~A., {Okudaira}, K., {Ohsumi}, K., {Ohnishi}, I.,
  {Nakamura-Messenger}, K., {Nakamura}, T., {Mostefaoui}, S., {Mikouchi}, T.,
  {Meibom}, A., {Matrajt}, G., {Marcus}, M.~A., {Leroux}, H., {Lemelle}, L.,
  {Le}, L., {Lanzirotti}, A., {Langenhorst}, F., {Krot}, A.~N., {Keller},
  L.~P., {Kearsley}, A.~T., {Joswiak}, D., {Jacob}, D., {Ishii}, H., {Harvey},
  R., {Hagiya}, K., {Grossman}, L., {Grossman}, J.~N., {Graham}, G.~A.,
  {Gounelle}, M., {Gillet}, P., {Genge}, M.~J., {Flynn}, G., {Ferroir}, T.,
  {Fallon}, S., {Ebel}, D.~S., {Dai}, Z.~R., {Cordier}, P., {Clark}, B., {Chi},
  M., {Butterworth}, A.~L., {Brownlee}, D.~E., {Bridges}, J.~C., {Brennan}, S.,
  {Brearley}, A., {Bradley}, J.~P., {Bleuet}, P., {Bland}, P.~A., {Bastien},
  R., Dec. 2006. {Mineralogy and Petrology of Comet 81P/Wild 2 Nucleus
  Samples}. Science 314, 1735--.

\bibitem[{{Zsom} et~al.(2010){Zsom}, {Ormel}, {G{\"u}ttler}, {Blum}, and
  {Dullemond}}]{zsom_etal_2010}
{Zsom}, A., {Ormel}, C.~W., {G{\"u}ttler}, C., {Blum}, J., {Dullemond}, C.~P.,
  Apr. 2010. {The outcome of protoplanetary dust growth: pebbles, boulders, or
  planetesimals? II. Introducing the bouncing barrier}. A\&A 513, A57.

\end{thebibliography}

\section*{Acknowledgements}

\vspace{1cm} We acknowledge Ulysse Marboeuf, together with James Owen and Philippe Rousselot for scientific discussion. We also warmly thank 
Pierre Morel for useful advices concerning numerical aspects, and we are grateful to Panayotis Lavvas for his scientific comments and for
reading the manuscript. We express our grateful thanks to Jeff Cuzzi who provided us his opacity code.
Simulations have been executed on computers from the Utinam Institute of the Universit\'{e} de Franche-Comt\'{e}, supported by the 
R\'{e}gion de Franche-Comt\'{e} and Institut des Sciences de l'Univers (INSU). We thank S\'{e}kou Diakit\'{e} who help us to parallelize 
the \textit{E}v\textit{AD} source code and to deal with the UTINAM Institute cluster. 
 Finally, we thank the anonymous Reviewers who improved the clarity of
the paper with their remarks and comments.

\end{document}